\documentclass[prb, twocolumn,superscriptaddress]{revtex4-1}
\usepackage{amsmath,amssymb,bm}
\usepackage{hyperref}
\usepackage{graphicx}
\usepackage{epstopdf}
\usepackage{latexsym}
\usepackage{subfigure}
\usepackage[usenames, dvipsnames]{color}
\usepackage[usenames, dvipsnames]{xcolor}
\usepackage{natbib}
\usepackage{braket}
\usepackage{float}
\usepackage[normalem]{ulem}
\usepackage{comment}
\usepackage{mathtools}
\usepackage{array}
\usepackage{tabu}
\usepackage{multirow}

\newcommand\redsout{\bgroup\markoverwith{\textcolor{red}{\rule[0.5ex]{2pt}{0.4pt}}}\ULon}
\newcommand\bluesout{\bgroup\markoverwith{\textcolor{blue}{\rule[0.5ex]{2pt}{0.4pt}}}\ULon}

\newcommand{\ADhide}[1]{{}}
\newcommand{\SPhide}[1]{{}}

\newcommand{\ADEDITOKAY}[2]{{}{\textcolor{black}{#2}}}

\begin{document}
\title{On the topological character of three-dimensional Nexus triple point degeneracies}
\author{Ankur Das}
\affiliation{Department of Physics and Astronomy, University of 
Kentucky, Lexington, KY 40506, USA}
\affiliation{Department of Condensed Matter Physics, Weizmann
Institute of Science, Rehovot, 76100 Israel}
\author{Sumiran Pujari}
\affiliation{Department of Physics, Indian Institute of
Technology Bombay, Mumbai, MH 400076, India}
\begin{abstract}

Recently a generic class
of three-dimensional band structures was identified
that host two-fold line degeneracies meeting
at three-fold or triple point degeneracies, which resist
the usual topological characterization of 
isolated point degeneracies as in Dirac/Weyl semimetals. 
For these so-called ``Nexus" fermions which 
lie beyond Dirac/Weyl fermions, we lay
out several concepts to characterize the wavefunction geometry
and spell out its topology. Our approach is based on an
understanding of the analyticity properties of Nexus
wavefunctions building on a two-dimensional analogue studied
recently by us. 
We use this to write down a homological \ADEDITOKAY{classification}{characterization}
of various Nexus triple point degeneracies in three dimensions.
\end{abstract}
\maketitle


\section{Introduction}
\label{sec:intro}

Band theory of electronic structure occupies a venerable place in
quantum condensed matter. Historically,
the basic ideas were established quite soon after the development 
of quantum mechanics. Yet, it is still an active area of research 
with many surprises. Among the surprises,
topological band insulators and superconductors
have captured our imagination in a big way. \cite{Hasan_Kane_review_2010}
They had 
their antecedent in the integer quantum Hall effect.
\cite{TKNN_1982,Klitzing_Dorda_Pepper_1980} The electronic structure of these quantum states 
of matter have  interesting and robust phenomenology, e.g. the
edge states of topological bands. \cite{Wen_1991,Hasan_Kane_review_2010}
Another surprise has
been the wealth of physics present in two and three dimensional ($2d$ and $3d$)
semimetals. \cite{Vafek_Vishwanath_review_2014,
Armitage_Mele_Vishwanath_review_2018}
The low energy excitations in semimetals also often possess
a topological character. This can lead to a
certain robustness against back-scattering.
\cite{Wehling_Black-Schaffer_Balatsky_review_2014} Already the
low density of semimetallic carriers at the Fermi energy makes
the effect of interactions less relevant. The combination of
these two effects holds promise for technological
applications of semimetals. \cite{Yan_Felser_2017}

From a theoretical point of view, what gives the semimetallic
carriers their topological character is the global structure of
their wavefunction geometry in the Brillouin zone.
Dirac and Weyl semimetals are the well-known
examples in $2d$ and $3d$.
These semimetals have two-fold degeneracies (not counting spin) 
at isolated points in
the Brillouin zone often protected by certain symmetries.
\cite{Vafek_Vishwanath_review_2014,Das_Kaul_Murthy_2020}
They can be thought as ``topological defects" in the space
of the band wavefunctions.
The semimetallic character obtains when
the Fermi energy is near these degeneracies. Such two-fold
point degeneracies are generic only in $3d$,
while they are exceptional in $2d$ thus requiring 
symmetry protection. \cite{VonNeuman_Wigner_1929}
Recently, generalization of Dirac and Weyl fermions have also been 
found by symmetry-protecting higher-fold point degeneracies. 
\cite{Bradlyn_etal_2016}

While there has been tremendous activity on semimetals
with point degeneracies, it has also been realized
that band structures with two-fold line degeneracies are
another possibility in the universe of possible band structures. Line
degeneracies are exceptional in $3d$, and symmetry
protection is required to obtain them. Several symmetry protected
possibilities have been identified recently.
\cite{Burkov_Hook_Balents_2011,Phillips_Aji_2014,Zhu_etal_2016,Chang_etal_2017}
Among these, there is a class of band structures where two-fold
line degeneracies meet at three-fold or triple point degeneracies. 
They have been dubbed as Nexus fermions. \cite{Chang_etal_2017,Heikkila_Volovik_2015}
There have been material proposals 
\cite{Chang_etal_2017,Zhang_etal_2017,Feng_etal_2018,Weng_etal_2016,Weng_etal_2016_2,Hyart_Heikkila_2016}
and experimental observations \cite{Lv_etal_2017,Ma_etal_2018} on this class of fermions.
Their spectral structure is intriguing, and their band topology
has been analysed previously in terms 
of the line degeneracies and 
$\mathbb{Z}_2$ topological numbers. \cite{Winkler_Singh_Soluyanov_2019}
The goal of this paper is
to shed more light on the band topology of Nexus fermions
in a different manner which particularly emphasizes
Nexus triple points themselves.
We want to \ADEDITOKAY{classify}{characterize} the topology of these triple point
degeneracies when thought of as defects in the space of band
wavefunctions.

The topological character of point degeneracies
can be understood by studying the band topology 
in one lower dimension. \cite{Schnyder_etal_2008}
One generally considers a surface in the 
momentum space that encloses the $3d$ point degeneracy in 
question. Since the surface can be chosen to be gapped
everywhere, one then computes the Chern number on this surface
which serves as a topological charge for the point degeneracy.
This discrete topological charge can not be changed by small
deformations to the Hamiltonian. This approach will fail to
characterize a Nexus triple point degeneracy, because any surface
enclosing it will have gapless points where the line degeneracies
intersect with the chosen surface.  Thus the general principle of
calculating a topological charge on an enclosing surface will not
work. This is why Ref. \onlinecite{Chang_etal_2017} called 
Nexus fermions as ``beyond-Weyl".
If we restrict ourselves to use only gapped lower dimensional
spaces, one can \ADEDITOKAY{at best}{}
characterize the topology of the line degeneracies by considering
gapped loops around them. \cite{Heikkila_Volovik_2015,Zhu_etal_2016,Winkler_Singh_Soluyanov_2019}
Also, one can enclose two or more
different triple points together by a gapped surface 
in some cases and calculate a topological
invariant on that surface. \cite{Zhu_etal_2016}

The question then is how to proceed in order to
characterize the band topology
of a Nexus system. This includes the basic issue
of whether Nexus triple points have a topological character or not.
This is a relevant question not just as a conceptual issue,
but also because of the following physical point: 
in Weyl systems, the surface Fermi arcs have a protection
in the sense that they have to end at the projection of the bulk Weyl
points on to the surface. \cite{Armitage_Mele_Vishwanath_review_2018}
This protection
is linked to the fact that the Weyl points in the bulk possess
a topological character. Ref. \onlinecite{Chang_etal_2017} raised
the analogous question on whether the surface Fermi arcs
numerically observed in their chosen Nexus systems
have a topological protection in the sense of Weyl Fermi arcs.
See the discussion on the
Nexus Fermi arcs in Ref. \onlinecite{Chang_etal_2017} for more
on this point.
Our paper gives a constructive
method to capture the topological character of different
Nexus triple points.
This method is the main result of this paper. 
Thus, we give an affirmative answer to
the question raised in Ref. \onlinecite{Chang_etal_2017}, i.e.
there will be surface Fermi arcs in Nexus systems that will
have to end at the projection of the bulk triple points on to
surface. 
We note here that Ref. \onlinecite{Zhu_etal_2016}
give an alternate argument for the presence of protected Fermi
arcs in these systems based on mirror Chern numbers
\cite{Teo_Fu_Kane_2008} without concerning directly with the
\ADEDITOKAY{classification}{topological character} of the Nexus points.

Our method relies crucially on
the analytic properties of the band wavefunctions near the line
degeneracies. This builds on the results of Ref.
\onlinecite{Das_Pujari_2019} where a toy $2d$ band structure
was considered which had a certain likeness to the Nexus band 
structures. In particular, specific $2d$ cuts of some Nexus
band structure resembles the toy band structure considered in
Ref. \onlinecite{Das_Pujari_2019}. The wavefunctions of this toy
model were written down which made the band topology
explicit. The $2d$ topology could be captured
by a generalization of winding numbers.
\cite{Park_Marzari_2011,Das_Pujari_2019}
This taught us the bigger lesson that near
line degeneracies, analytic continuation or movement in the space
of wavefunctions is key to exposing the band topology
even in $3d$.

Motivated by the above, we will study in detail the analyticity 
properties of several $3d$ Nexus band structures. 
We will use Dirac and Weyl systems as scaffolding for the
analyticity discussions of Nexus band structures. In the process,
we will come to an important notion of the \emph{generalized} 
domain when dealing with degeneracies. This will be
necessitated by the presence of
degenerate points on the surface enclosing the 
Nexus triple point degeneracy. For point
degeneracies like Weyl points, this notion is not necessitated
because we can easily find a gapped surface to surround the Weyl
point.

Equipped with the generalized domain, we can finally
state data on the band topology of a Nexus band structure.
This scheme will consist
of specifying and counting the distinct analytic loops that can
be drawn on the generalized domain around a triple point.
Thus we will have the desired scheme to distinguish different
triple point degeneracies based on their distinct band topology
data.  This idea is very similar to the homology classes of
1-cycles used to distinguish the topology of different geometric
objects. \cite{Nakahara_2003} The familiar example is that of a
sphere vs. a torus.
The sphere admits no loops that can't be contracted to a point,
whereas a torus admits two distinct classes of loops that can't
be contracted to a point. Our scheme will do a similar
\ADEDITOKAY{classification}{characterization} of the triple points, with the structure 
of the homology classes being dictated by the structure of the line
degeneracies. In this way, we will be able to \ADEDITOKAY{classify}{describe} several
Nexus band structures written down in the literature
\cite{Chang_etal_2017} as well as some obtained as $3d$
extensions of the toy band structure in Ref.
\onlinecite{Das_Pujari_2019}. This \ADEDITOKAY{classification}{} is the 
culminating result of this paper.
Furthermore, this scheme can also potentially reveal the
inter-relationships between different kinds of triple points.

We give a brief outline of the paper: Sec.
\ref{sec:2d_analyticity} sets the stage by recapitulating some
$2d$ band structures from the point of view
of analyticity. We will be paying close attention to
what happens near degeneracies, since that is the main roadblock
in understanding the band topology of Nexus band structures.
Doing this will introduce the notion of the generalized
domain. We then go on $3d$
in Sec. \ref{sec:3d_analyticity}. We start by discussing the
familiar Weyl system to give a clear contrast to Nexus band
structures in terms of their analyticity properties. We then discuss
several Nexus band structures.
Sec. \ref{sec:classification} will
finally give the method to state the band topology data in terms
of homology classes of analytic loops on a generalized
domain around the triple point. This won't be hindered
by a lack of gapped property, because analyticity near the
degeneracies constrain the wavefunctions enough to enable
stating the topology. This will conclude our exposition on the
band topology of Nexus fermions. We end the paper in Sec.
\ref{sec:conclusion} with a summary and outlook.
We also discuss here our take on the Fermi arc
phenomenology of Nexus systems including a conjecture regarding
the charge of these surface states.


\section{2D Analyticity}
\label{sec:2d_analyticity}

In this section, we will start with the analyticity discussion
in a $2d$ beyond-Dirac Nexus system.
Let's reconsider the band structure introduced in 
Ref. \onlinecite{Das_Pujari_2019} to set up the discussion: 
\begin{equation}
    H(\mathbf{p})=\left(\begin{matrix} 
            0 & p_x-i p_y & p_x-ip_y \\
            p_x+i p_y & 0 & p_x+ip_y\\
            p_x+ip_y & p_x-ip_y & 0
           \end{matrix}\right)
        \label{eq:3bandcontham}
\end{equation}
The eigensystem of $H(\mathbf{p})$ is
\begin{subequations}
\begin{align}
    \epsilon_\alpha (\mathbf{p}) & = 2 p \cos\left[ \frac{\theta_\mathbf{p}}{3} + (2+\alpha)\frac{2\pi}{3}\right] \\
    v_\alpha(\mathbf{p}) & =\frac{1}{\sqrt{3}}\left( \omega^{2+\alpha} e^{-i\frac{2\theta_\mathbf{p}}{3}}~~~
    (\omega^*)^{2+\alpha}
    e^{i\frac{2\theta_\mathbf{p}}{3}}~~~1 \right)^T
\end{align}
\label{eq:eisys_3band}
\end{subequations}
where $\theta_{\mathbf{p}} = \arctan \left(\frac{p_y}{p_x} 
\right) \in [0,2\pi)$. $\omega = e^{i \frac{2\pi}{3}}$, 
$\omega^2= e^{-i \frac{2\pi}{3}}$ are the complex cube roots of 
unity and $\alpha=0,1,2$. This band structure possesses a
three-fold degeneracy at $\mathbf{p}=0$ clearly,
and has two line degeneracies coming out from the triple point 
which is a signature feature of Nexus wavefunctions. Because of
the line degeneracies, a standard Berry phase description of the
wavefunction geometry is not applicable. However, we had used
generalized winding numbers
\cite{Das_Pujari_2019,Park_Marzari_2011} to understand
this $2d$ wavefunction geometry (cf. Table I and Sec. II of Ref. 
\onlinecite{Das_Pujari_2019}) and contrasted with other known
$2d$ Dirac-like wavefunction geometries. In $3d$, such winding
number description is not generally applicable for classification
of point degeneracies. Thus, we will 
take the approach to 
be described below and in future sections.

Our main point of view
will be to understand and write down the key aspects of the 
analytic behavior of various band structures.
This is a different way of communicating invariant
data of the wavefunction geometry
than winding numbers and Berry phases. For example, we often view
the familiar two-fold Dirac system 
\begin{align}
H_K^\text{Dirac}(\mathbf{p})=\left(\begin{matrix} 
            0 & p_x-i p_y \\
            p_x+i p_y & 0 
           \end{matrix}\right)
\label{eq:HamDirac}
\end{align}
with the eigensystem as
\begin{align}
\epsilon_\pm(\mathbf{p})= \pm p  \: ; \: & \: \: v_\pm(\mathbf{p})=\frac{1}{\sqrt{2}} \left( \pm e^{-i \theta_{\mathbf{p}}}, 1 \right)^T  
\label{eq:eigsys_dirac}
\end{align}
by calculating Berry phase or chiral winding number
\cite{Ryu_etal_2010,Schnyder_etal_2008} on gapped region in
one lower dimension (e.g.  any closed loop around the 
degeneracy). We rather want to include the degeneracy to be a
part of the analysis.

\begin{figure}[h]
    \centering
    \includegraphics[trim=150 200 150
    100,clip,width=0.227\columnwidth]{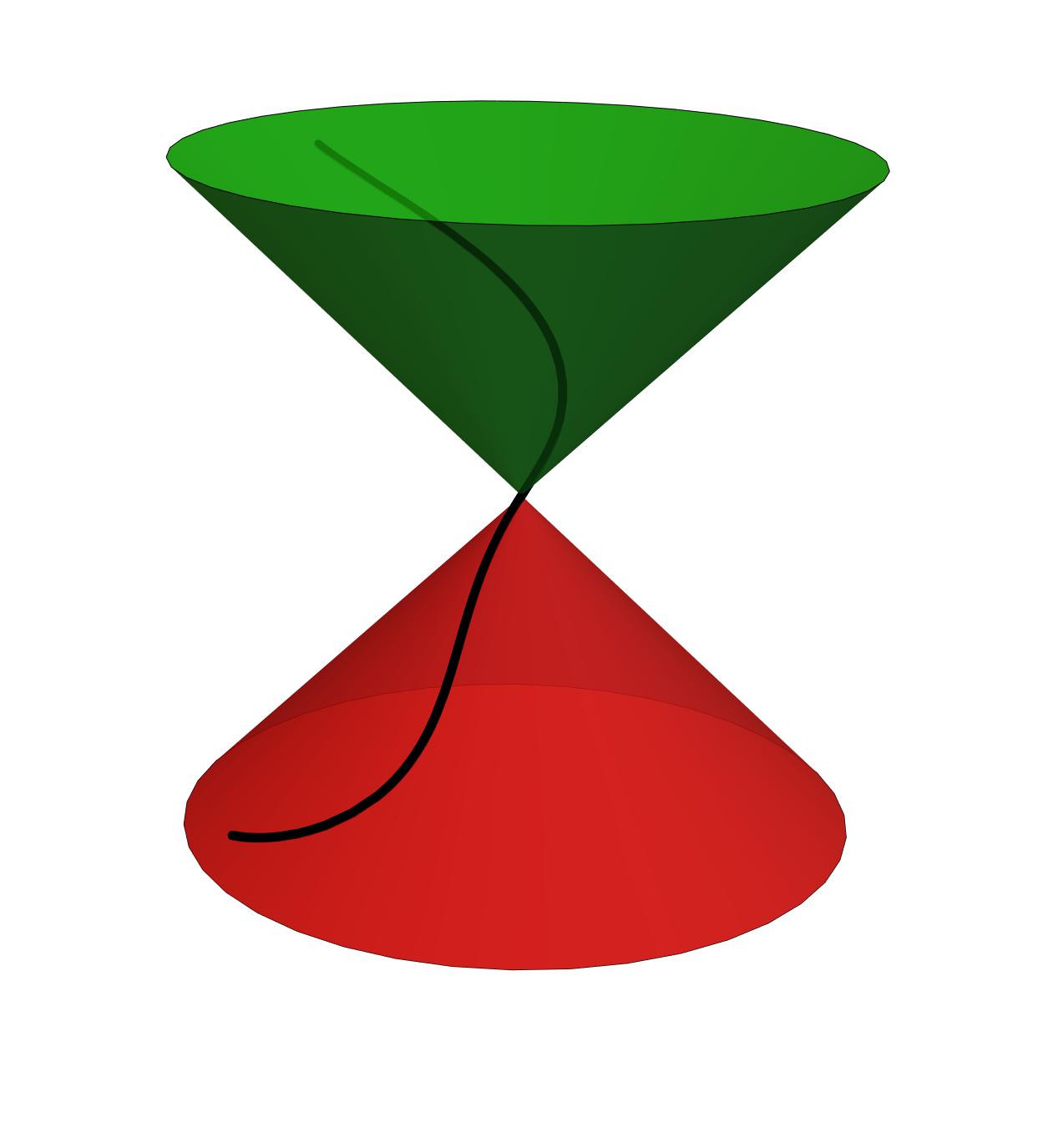}
    \includegraphics[trim=0 20 0
    20,clip,width=0.25\columnwidth]{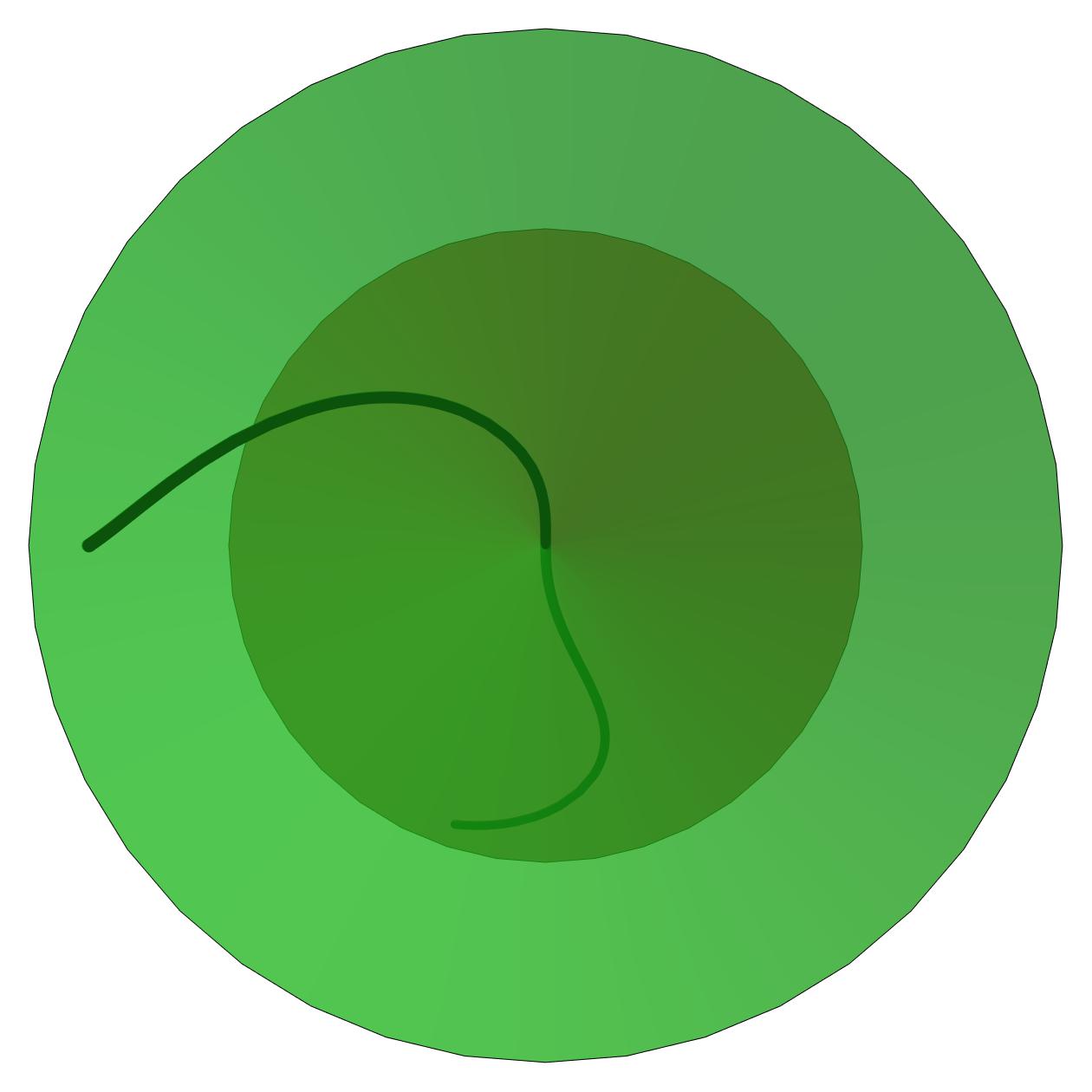}
    \includegraphics[trim=0 95 0 95,clip,width=0.3875\columnwidth]{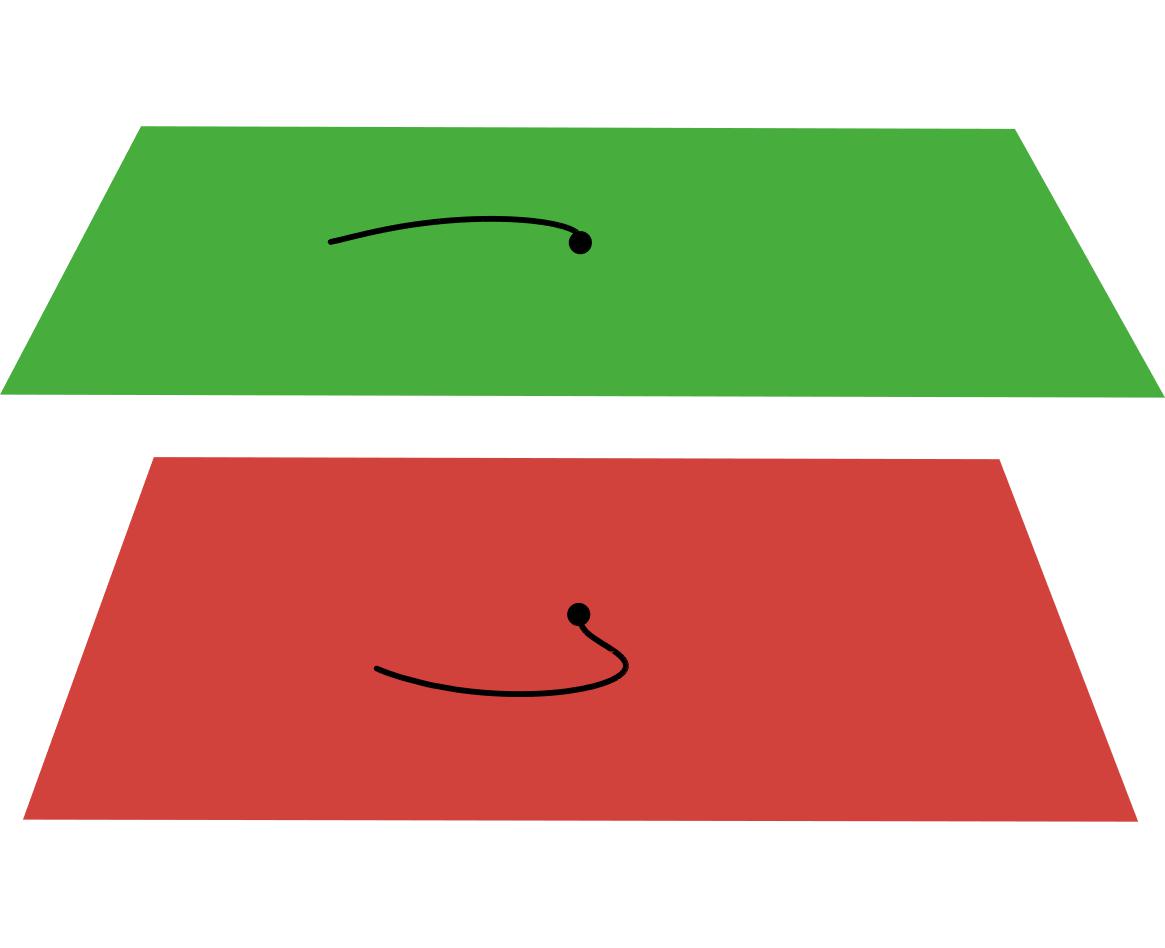}
    \caption{
    This figure illustrates the analytic way of moving
    across or through the Dirac point. The left and middle figures show
    how that analytic
    movement happens in the spectrum from the side and top
    views respectively. The right figure shows the analytic movement in the
    generalized domain. As mentioned in the
    text, the generalized domain  is made of
    two copies of $p_x$, $p_y$ plane connected at the 
    Dirac point. The color scheme is only for convenience.}
    \label{fig:dirac_move_sketch}
\end{figure}

Firstly, on a gapped loop we clearly have the analyticity property
\begin{align}
    v_i(\theta+2\pi)=v_i(\theta)
    \label{eq:dirac_analyticity_bp}
\end{align} 
However, we also have the following
analyticity property of Dirac wavefunctions
\begin{align}
    v_+(\theta+\pi)=v_-(\theta)
    \label{eq:dirac_analyticity_bc}
\end{align} 
which connects the two bands. 
In fact, this relation tells us how to consistently arrive
at the two-fold degeneracy from \emph{all} sides without
running into analytic ambiguities.
Thus, we can interpret this as 
the way to move analytically \emph{across} the 
point degeneracy.
This is illustrated in the two figures from the left in
Fig. \ref{fig:dirac_move_sketch}.

\onecolumngrid

\twocolumngrid

Eq. \ref{eq:dirac_analyticity_bp} and \ref{eq:dirac_analyticity_bc}
are nothing but an alternate way of describing the wavefunction 
geometry that is captured by Berry phase and chiral winding
numbers, with the additional benefit of allowing to move
across the degeneracy in an analytically smooth
way. This alternate viewpoint will prove useful for us
because Nexus triple points can not be enclosed
by a gapped region in one lower dimension.
As notation, we refer to analyticity relations with the same
band index on left and right hand sides as ``index-preserving"
(e.g. Eq. \ref{eq:dirac_analyticity_bp}), 
while analyticity relations with different band indices on both
sides as ``index-connecting" (e.g. Eq.
\ref{eq:dirac_analyticity_bc}).

\begin{figure}[ht]
    \centering
    \includegraphics[trim=0 530 0 600,clip,width=0.8\columnwidth]{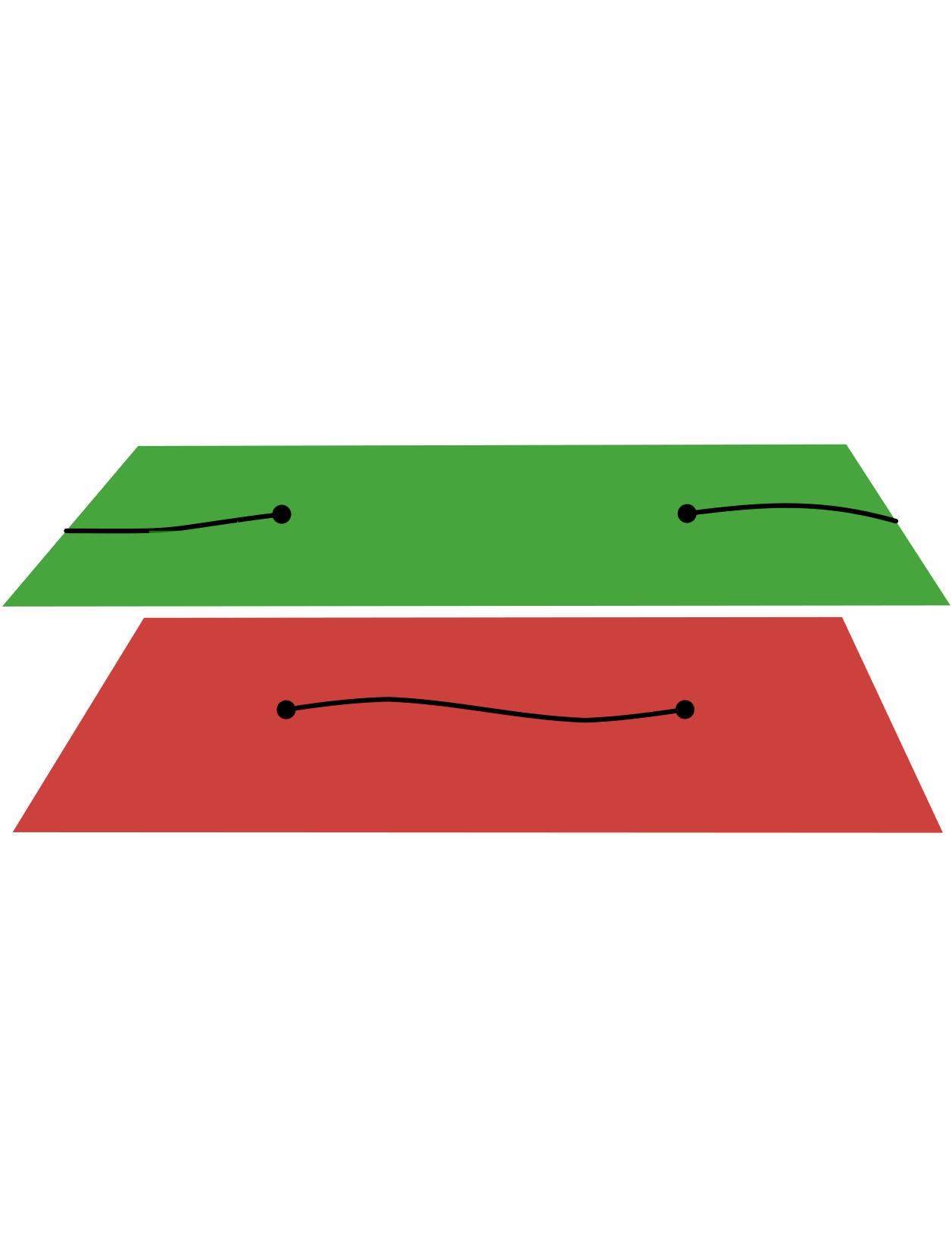}
    \vspace*{0.5cm}
    
    \includegraphics[trim=0 530 0 600,clip,width=0.8\columnwidth]{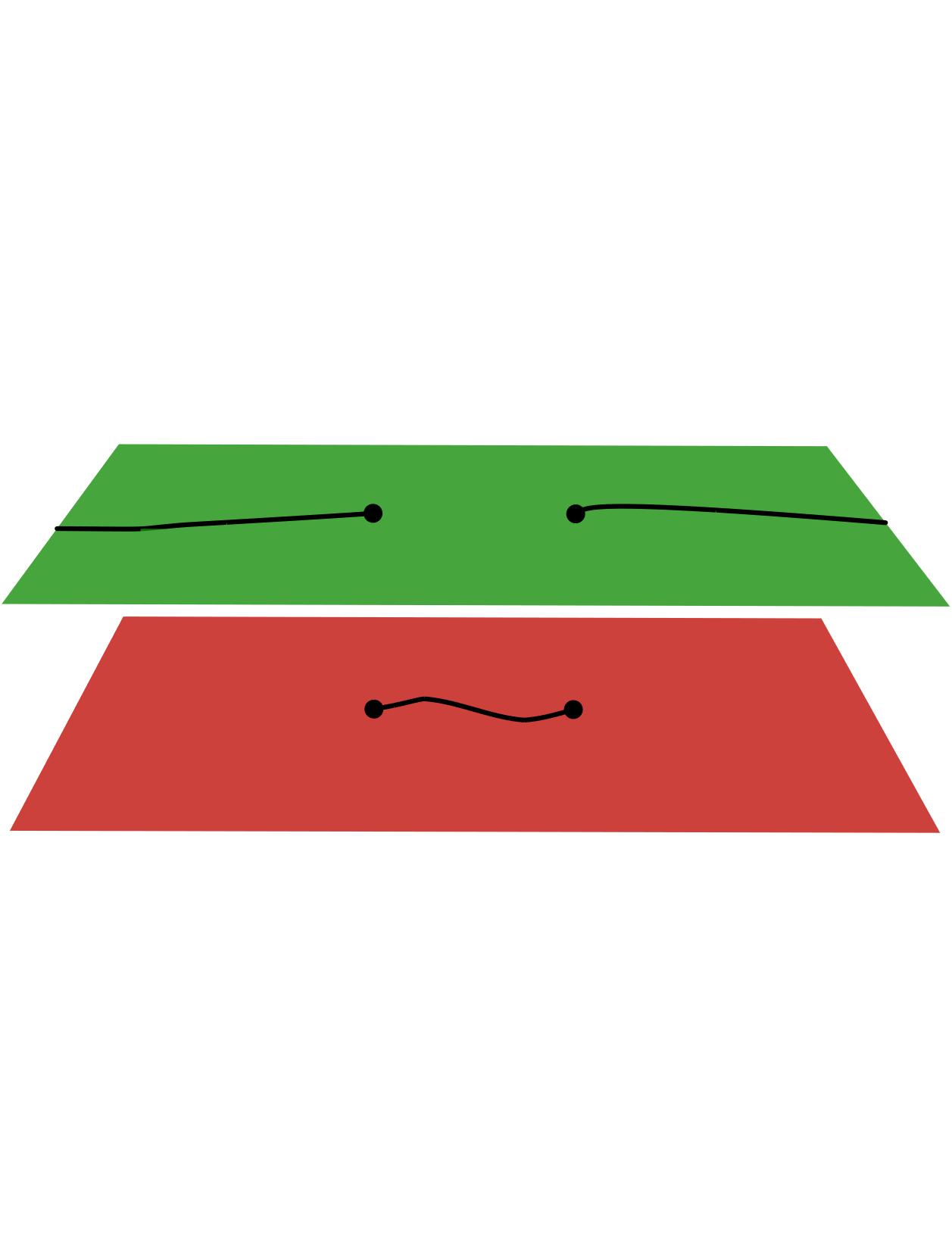}
    \vspace*{0.5cm}
    
    \includegraphics[trim=0 530 0 600,clip,width=0.8\columnwidth]{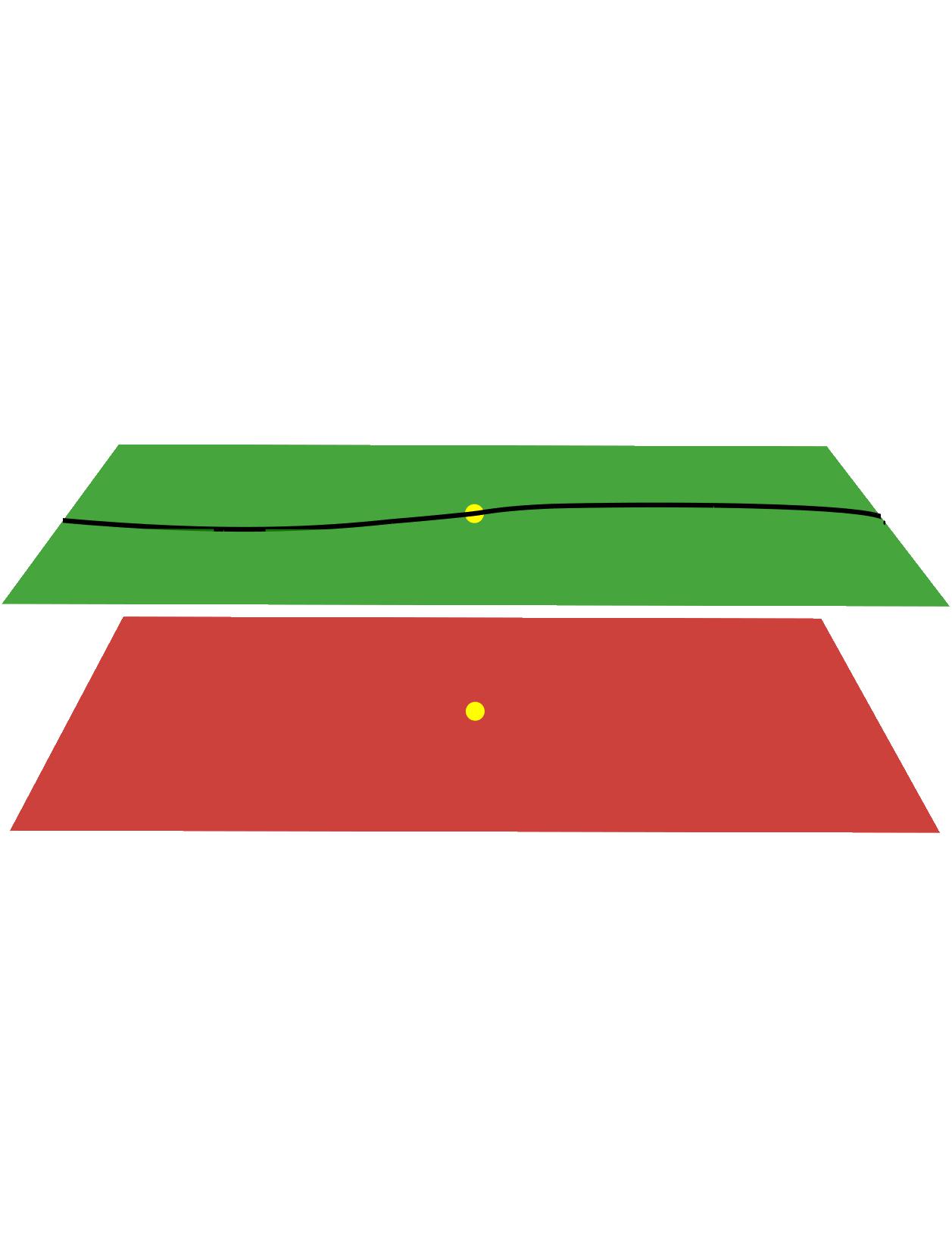}
    \caption{An illustration of the analytic movement in the case
    of a quadratic band touching (QBT). The top two panels
    show how the analytic behavior of a pair of Dirac points 
    reduces to that of a QBT in the last panel. 
    Thus we see how a single Dirac touching and a QBT differ in 
    their analytic movements across the degenerate point.}
    \label{fig:qbt_move_sketch}
\end{figure}

For a quadratic band touching (QBT), the analyticity relation is
in fact 
\begin{align}
    v_+(\theta+\pi)=v_+(\theta).
    \label{eq:qbt_analyticity_bc}
\end{align} 
We can understand this in terms of two Dirac points (of same
winding) sitting on top of each other. Let's first imagine these
two Dirac points are not on top of each other,
and we move analytically across both the degeneracies in a single
go. In this process, we will return back to the same band that
we started from as illustrated in Fig. \ref{fig:qbt_move_sketch}.
Now, 
imagine moving these two Dirac points till they fall on top of
each other to obtain a QBT. Analyticity thus
forces us that we will stay in the same band when we cross the
QBT (bottom panel of Fig.
\ref{fig:qbt_move_sketch}), i.e.     Eq. \ref{eq:qbt_analyticity_bc}.
This argument also works when the QBT splits
into more Dirac points, e.g. in Bernal-stacked honeycomb bilayer
lattice in presence of ``trigonal" warping terms when it 
splits into three Dirac points of same winding and a fourth one
with opposite winding. \cite{Mikitik_Sharlai_2008}

With the above discussion in hand, we can revisit the 
$2d$ system in Eq. \ref{eq:3bandcontham} and 
\ref{eq:eisys_3band} in terms of its key analyticity information.
The index-connecting relation is
\begin{align}
    v_{\alpha}(\theta+2\pi)=
v_{(\alpha+1) \text{ mod } 3}(\theta)
\label{eq:2dnexus_analyticity_bc}
\end{align}
which may be contrasted with Eq. \ref{eq:dirac_analyticity_bc} in
the Dirac case. Similarly, in contrast to Eq. 
\ref{eq:dirac_analyticity_bp}, we arrive at the index-preserving
relation
\begin{align}
    v_{i}(\theta+6\pi)= v_{i}(\theta)
\label{eq:2dnexus_analyticity_bp}
\end{align}

This unusual ``$+ 6\pi$" structure is a result of the Nexus lines
emanating from the three-fold degeneracy. We emphasize that the above two
relations are an alternate way of describing the wavefunction
geometry when compared to a generalized winding number
description. \cite{Das_Pujari_2019} This way of
stating the wavefunction geometry via the analyticity
will be our approach to tackle the $3d$ Nexus 
geometry in the next sections.

We end this section with a final conceptual point. Even though
Eq. \ref{eq:3bandcontham} and \ref{eq:HamDirac} are multi-band
systems as expressed through band-indexed eigenfunctions in Eq.
\ref{eq:eisys_3band} and \ref{eq:eigsys_dirac}, the analytic
structure actually tells us that this band distinction is a
matter of convenience or convention 
and not fundamental when considering the
wavefunction geometry. We can imagine a single function 
defined on a \emph{generalized domain} that describes the
multi-band wavefunctions in analogy with Riemann surfaces. This
analogy can be made exact for Eq. \ref{eq:eisys_3band} by
re-writing as
    $\epsilon_\alpha (\mathbf{p})=  2 p ~ \text{Re}\left[ \omega^{2+\alpha} e^{i\frac{\theta_\mathbf{p}}{3}}\right]$,
    $v_\alpha(\mathbf{p}) =\frac{1}{\sqrt{3}}\left( \omega^{2+\alpha} e^{-i\frac{2\theta_\mathbf{p}}{3}}~~~
    (\omega^*)^{2+\alpha}
    e^{i\frac{2\theta_\mathbf{p}}{3}}~~~1 \right)^T$
whereby we can essentially drop the $\alpha$ index to write 
as  $\epsilon (\mathbf{p})=  2 p ~ \text{Re}\left[ e^{i\frac{\theta_\mathbf{p}}{3}}\right]$,
    $v(\mathbf{p}) =\frac{1}{\sqrt{3}}\left(  e^{-i\frac{2\theta_\mathbf{p}}{3}}~~~
    e^{i\frac{2\theta_\mathbf{p}}{3}}~~~1 \right)^T$
that is defined on a three-fold Riemann surface connected
by branch cuts of the complex cube root function.
This generalized domain restatement succinctly tells us how to
analytically move in the space of wavefunctions,
which is of course a key requirement to understand the
wavefunction geometry.

A similar generalized domain restatement can be done
for the case of Dirac eigensystem Eq. \ref{eq:eigsys_dirac}. 
The generalized domain is composed of \emph{two copies} 
of the $p_x$-$p_y$ plane connected at the point-degeneracy.
The band-connection relation (Eq. \ref{eq:dirac_analyticity_bc})
gives us the rule of moving through the ``connecting point" in the
generalized domain from one copy of the $p_x$-$p_y$ plane
to the other (see the rightmost figure of Fig. \ref{fig:dirac_move_sketch}). 
In the Dirac case, there is no branch cut 
structure since the eigensystem (Eq. \ref{eq:eigsys_dirac}) is 
perfectly analytic.
The generalized domain will be used when we discuss the
 $3d$ Nexus wavefunctions in the next
sections.


\section{3d Analyticity}
\label{sec:3d_analyticity}

In this section, we start with the actual discussion on the
analyticity properties of $3d$ Nexus fermions.
As mentioned in Sec. \ref{sec:intro}, 
line degeneracies are exceptional in $3d$
and require symmetry protection, whereas they are 
fine-tuned in $2d$.
Thus the analyticity discussion in the
previous section is for a fine-tuned case, 
but it will help us in the
following discussions. Before we go towards Nexus analyticity
properties, let us start with the familiar case of Weyl point
degeneracies to set the stage.

\subsection{Weyl analyticity}

A Weyl point degeneracy is characterized by an effective
(low-energy) Hamiltonian of the form $H^\text{Weyl} = \sum_{i \in \{x,y,z\}}
p_i \sigma_i$. The eigenenergies are $\epsilon(\mathbf{p}) = \pm
p$, and the eigenfunctions are generally expressed as
\begin{subequations}
\begin{align}
    v_+(\mathbf{p}) & =\left(e^{-i \phi} \cos(\theta/2)  ~~~\sin(\theta/2)\right)^T \\
v_-(\mathbf{p}) & =\left(- \sin(\theta/2) ~~~ e^{i \phi} \cos(\theta/2)\right)^T.
\end{align}
\label{eq:eigsys_weyl1}
\end{subequations}
In our gauge choice where the last term is kept purely real, they are
\begin{subequations}
\begin{align}
    v_+(\mathbf{p}) & =\left(e^{-i \phi} \cos(\theta/2)  ~~~\sin(\theta/2)\right)^T \\
v_-(\mathbf{p}) & =\left(- e^{-i \phi} \sin(\theta/2) ~~~  \cos(\theta/2)\right)^T.
\end{align}
\label{eq:eigsys_weyl2}
\end{subequations}
Often, the wavefunction geometry of $3d$
point degeneracies are understood by considering a 
$2d$ surface enclosing the point degeneracy and computing the 
Chern number of the two \emph{gapped} bands on this reduced 
$2d$ system. For the Weyl system, the Chern number of the two gapped
bands are $\pm 1$. We note that in the full BZ, the number of Weyl
points has to be even such that the sum of their
Chern numbers is zero, as the Chern number computed on the BZ
boundary must be zero by periodicity.

Another perspective on the Weyl geometry is the following
\onlinecite{Vishwanath_youtube_2015}: consider $2d$ cross-sections in
the Brillouin zone away from the point degeneracy, e.g. a
constant $k_z$ plane which is a representative $2d$ system.
In such cross-sections, we obtain a gapped Dirac cone 
system with the specific sign of the mass term controlled by
the sign of $p_z$. Because the $2d$ system is gapped, we may
compute a Chern number.
On either side of the Weyl point, the sign
of the mass changes. Thus, the Weyl degeneracy may be interpreted
as a transition between the two topologically different $2d$
Chern bands on either side.

However, anticipating the lack of gapped $2d$ surfaces in 
presence of line degeneracies for Nexus fermions, we may ask
what happens if we were to consider cross-sections which
\emph{always} include the Weyl point, e.g. consider any plane
going through the Weyl point. In particular, if we consider a
family of such planes -- e.g. all planes containing $p_z$-axis--,
then we would like to ask how does this family of $2d$ bands
interpolate among each other? This forces us to grapple with the
role of the degeneracy in the analysis. This is a similar
motivation to what we have done in $2d$ as in
Sec. \ref{sec:2d_analyticity} where 
stating the index-connecting relation is our way of answering 
this question.
In $3d$ we will need to make a choice of the
coordinate system, however for the Weyl discussion, the spherical
symmetry comes to our rescue and we can use the $p_z$ axis to set
up our spherical coordinates without any loss of generality.
The analyticity relations are the following:
\begin{subequations}
\begin{align}
    v_+(\pi-\theta,\phi+\pi) & = v_-(\theta,\phi)
    \label{eq:weyl_analyticity_bc} \\
    v_i(\theta,\phi+2\pi) & = v_i(\theta,\phi).
    \label{eq:weyl_analyticity_bp}
\end{align}
\end{subequations}
Graphically speaking, we
have to exit in the same ``direction" that we came in towards
the degeneracy. This is the exact same behavior as shown in Fig. \ref{fig:dirac_move_sketch}
in one higher dimension. We notice here that Eq.
\ref{eq:weyl_analyticity_bc} conveys the same information as the
changing sign of mass \cite{Vishwanath_youtube_2015}
in a different way. Finally, the generalized domain restatement
will now consist of two copies of the $p_x$-$p_y$-$p_z$ space
joined again at the point degeneracy with the above analyticity
relations (Eq. \ref{eq:weyl_analyticity_bc}, 
\ref{eq:weyl_analyticity_bp}) as the rules to move in this
generalized domain.

\subsection{Nexus analyticity}
\label{subsec:nexus_anal}

\begin{figure}[h]
    \centering
    \includegraphics[trim=150 150 50 0,clip,width=0.8\columnwidth]{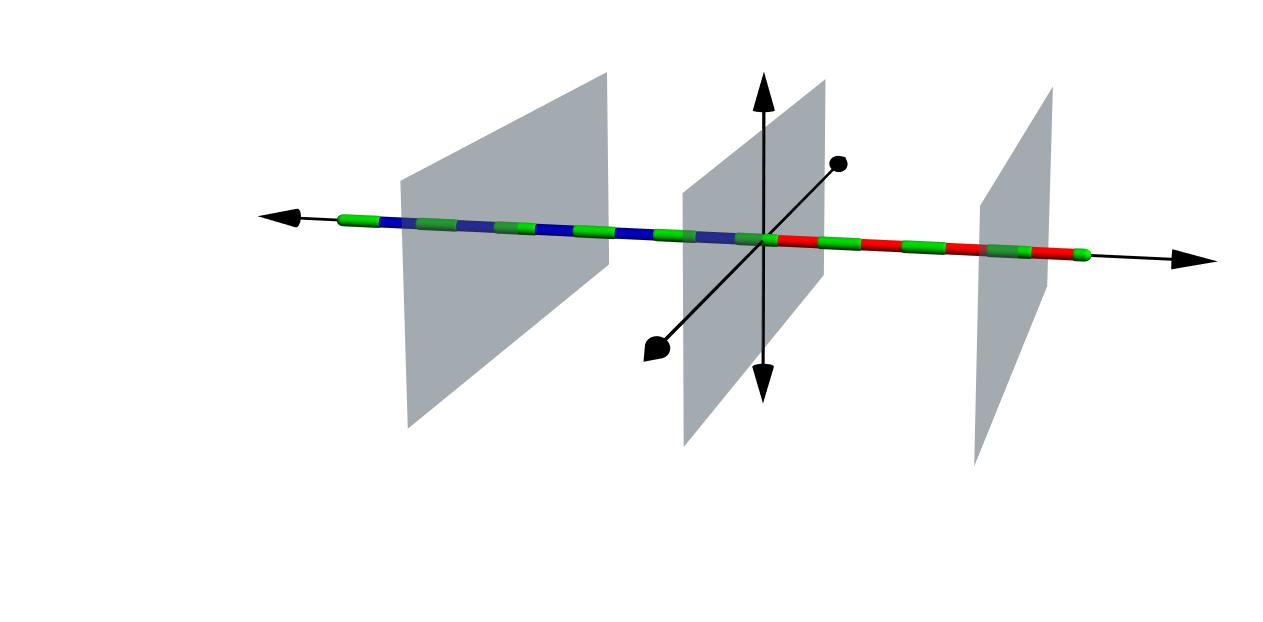}
    \setlength{\unitlength}{0.84cm}
    \put(0.6,1.6){\llap{\large $p_x$}}
    \put(-2.3,2.5){\llap{\large $p_y$}}
    \put(-3.25,3.2){\llap{\large $p_z$}}
    \put(-1.65,0.4){\llap{\large $\boxed{\epsilon_1}$}}
    \put(-3.65,0.4){\llap{\large $\boxed{\epsilon_2}$}}
    \put(-5.65,0.4){\llap{\large $\boxed{\epsilon_3}$}}
    
    \includegraphics[trim=0 180 0 150,clip,width=0.8\columnwidth]{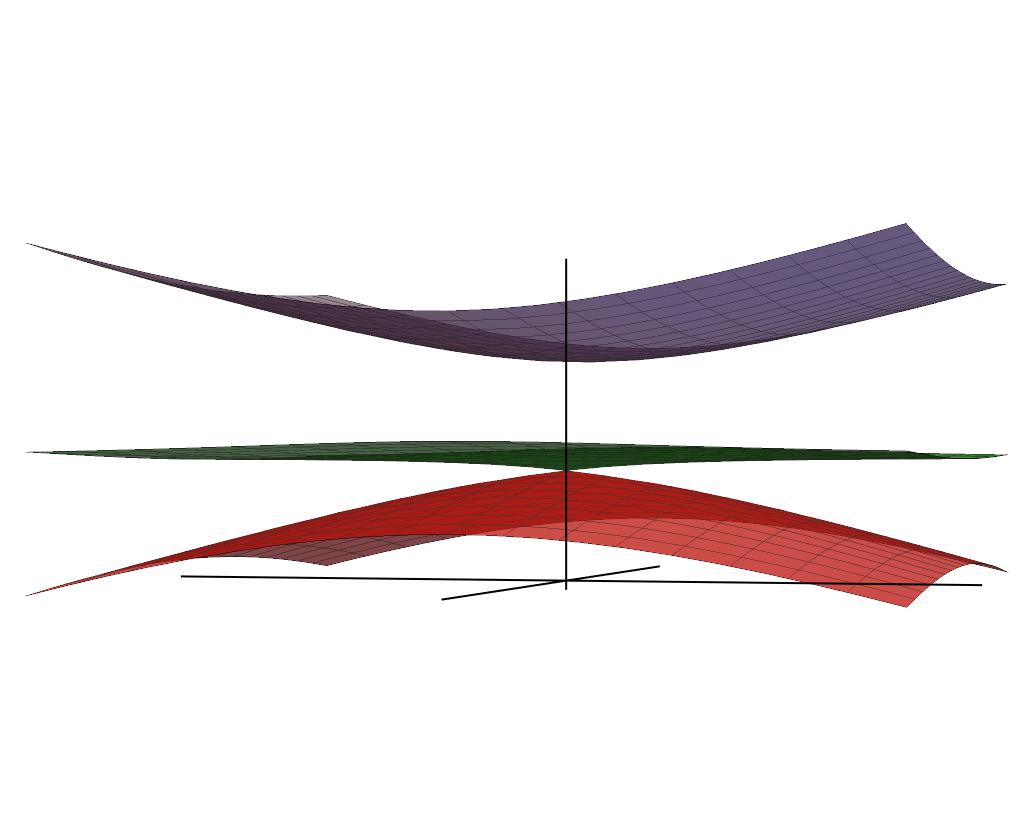}
    \setlength{\unitlength}{0.84cm}
    \put(0.25,0.3){\llap{\large $p_x$}}
    \put(-4.5,0.2){\llap{\large $p_z$}}
    \put(-3.5,3.25){\llap{\large $\epsilon_1$}}
    
    \includegraphics[trim=0 180 0 150,clip,width=0.8\columnwidth]{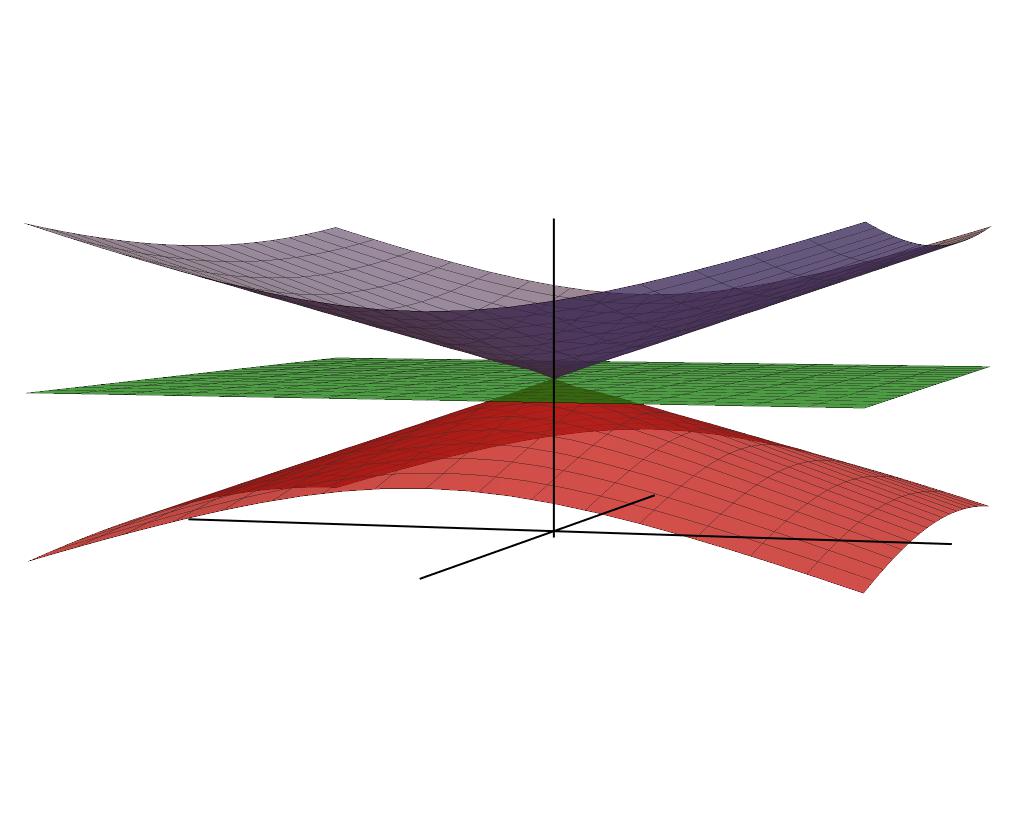}
    \setlength{\unitlength}{0.84cm}
    \put(0.,0.65){\llap{\large $p_x$}}
    \put(-4.75,0.3){\llap{\large $p_z$}}
    \put(-3.5,3.5){\llap{\large $\epsilon_2$}}
    
    \includegraphics[trim=0 180 0 150,clip,width=0.8\columnwidth]{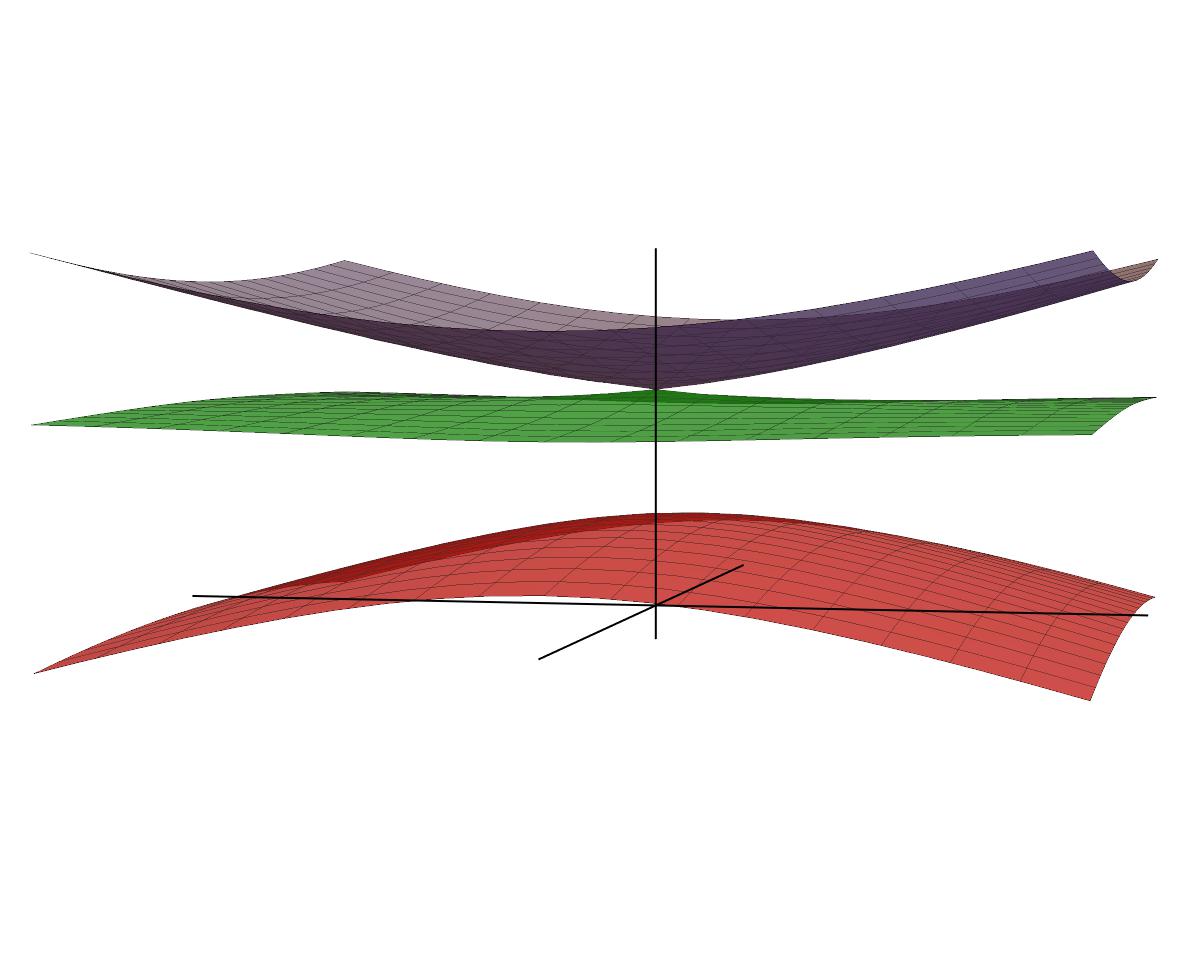}
    \setlength{\unitlength}{0.84cm}
    \put(0.3,1.0){\llap{\large $p_x$}}
    \put(-4.5,0.6){\llap{\large $p_z$}}
    \put(-3.5,3.8){\llap{\large $\epsilon_3$}}
    
    \caption{The top panel shows the line degeneracies
    as dashed alternate colored lines for $H^3$. 
    Red-Green-Blue stand for bottom, middle,
    and top bands respectively. 
    The following three panels show the band structure for 
    generic $2d$ cuts that intersect the line degeneracies
    highlighted in the topmost panel.
    }
    \label{fig:ham_lambda3z}
\end{figure}
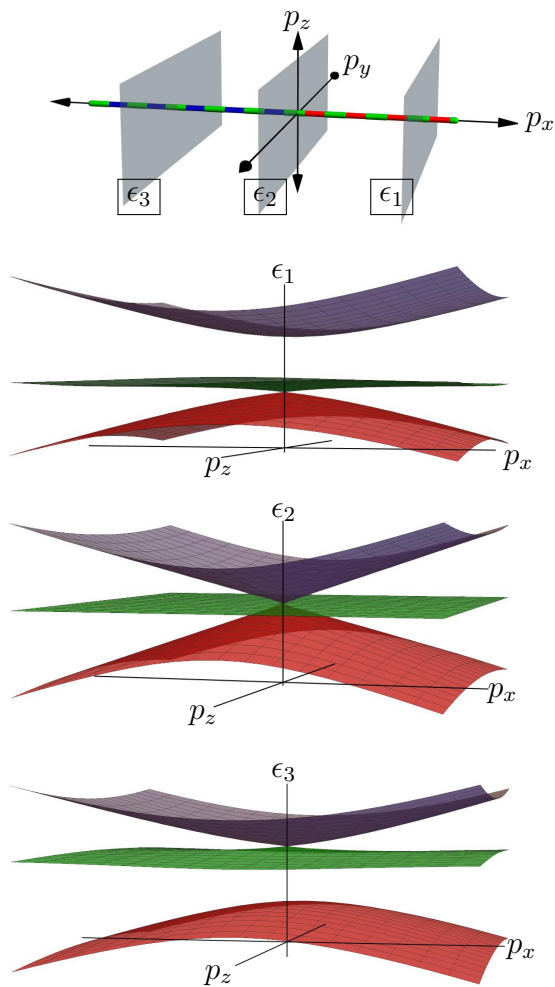

Now, we tackle the main case of $3d$ Nexus triple points.
Using $SU(3)$ generators $\Lambda^i$ (the Gell-Mann matrices \cite{Halzen_Martin_1984})
for brevity,
the $2d$ Nexus system (Eq. \ref{eq:3bandcontham}) looks like
\begin{equation}
    H(\mathbf{p})= p_x(\Lambda^1+\Lambda^4+\Lambda^6)+p_y(\Lambda^2+\Lambda^5-\Lambda^7).
\end{equation}
To this, we start by adding a diagonal $\Lambda^3$  ``mass" term 
linear in $p_z$ (in analogy with $p_z \sigma_z$ for the Weyl case)
such that we get a $3d$ Nexus triple point. 
Thus we have
\begin{equation}
    H^3(\mathbf{p})=H(\mathbf{p})+p_z \Lambda^3 \label{eq:ham_lambda3z}
\end{equation}
Fig. \ref{fig:ham_lambda3z} shows
the line-degeneracy structure and the triple point
given by Eq. \ref{eq:ham_lambda3z}.

Similar to the Weyl discussion, we will discuss 1) how do the 
(generic) $2d$ cross-sections away from the triple point evolve
as we cross the triple point, \cite{Winkler_Singh_Soluyanov_2019} and 2) what are the analyticity
relations that characterize the presence of triple points. 
We will sometimes refer to them as topological defects
or monopoles in analogy with Weyl point degeneracies (Sec.
\ref{sec:classification} will give a topological \ADEDITOKAY{classification}{characterization}
of these defects). Also, line degeneracies are extended
topological defects present in the Nexus system.
\cite{Heikkila_Volovik_2015} Ref. \onlinecite{Heikkila_Volovik_2015}
gave a $Z_2$ topological charge to the line degeneracy by
computing a $Z_2$ topological invariant (cf. Eq. 1 in
Ref. \onlinecite{Heikkila_Volovik_2015}) on a $d-2=1$ dimensional
loop around the line degeneracy.
One can also compute a chiral winding number \cite{footnote_winding} on such loops which
is a $Z$ invariant. \cite{Ryu_etal_2010,Schnyder_etal_2008,Faluga_Hassler_Akhmerov_2012}

The eigensystem formula for $H^3$ is comparatively involved than
Weyl eigensystem (Eq. \ref{eq:eigsys_weyl2}) and we do not write it
down explicitly. 
The exact details are not relevant to
understand the analyticity properties. Fig.
\ref{fig:ham_lambda3z} shows the evolution of (generic) $2d$ cuts
across the triple point. We see that on one side the top and
middle bands are joined by a Dirac point with the bottom band 
as standalone, while on the other side
the bottom and middle bands are joined by a Dirac point with
the top band as standalone.
The triple point is thus to be thought as a defect which
separates these two different behaviors.
We can think of these behaviors as two different $SU(2)$ groups,
\cite{Ramond_2010} one involving 
middle and top bands and another involving middle and
bottom bands. In comparison to the Weyl degeneracy, where the sign
of the Dirac mass changes on either side, here the triple point
degeneracy is changing one type of $SU(2)$ \emph{defect} to the
other type.

\begin{figure}[h]
    \centering
    \includegraphics[trim=150 0 50 100,clip,width=0.8\columnwidth]{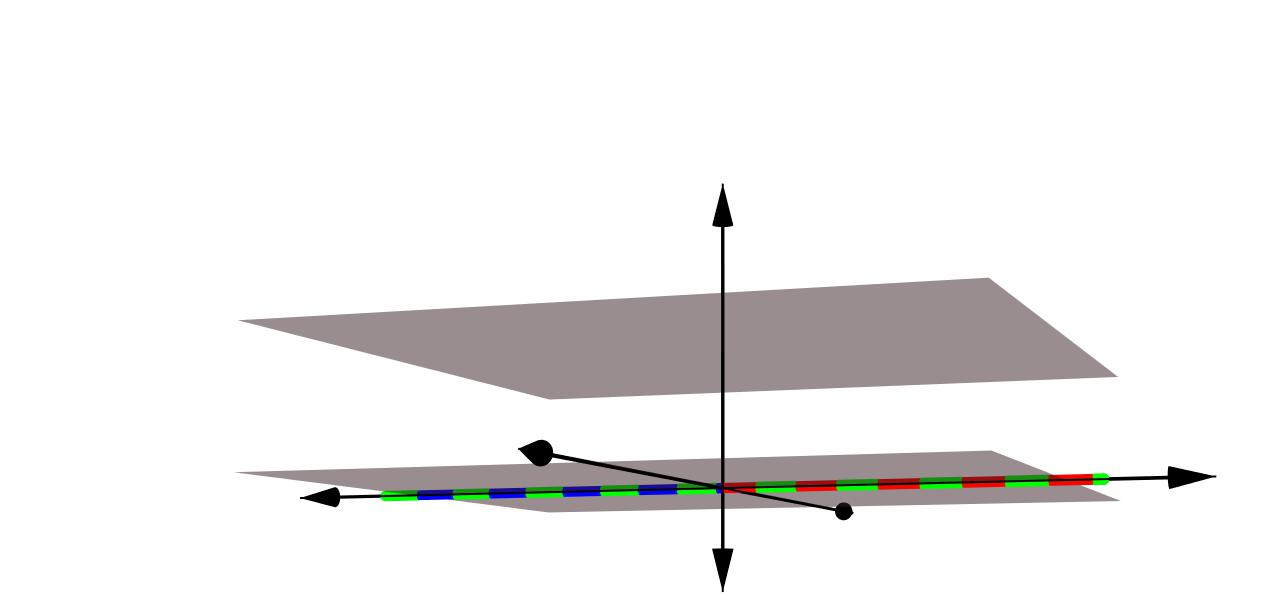}
    \setlength{\unitlength}{0.84cm}
    \put(0.6,0.9){\llap{\large $p_x$}}
    \put(-2.25,.45){\llap{\large $p_y$}}
    \put(-3.6,3.5){\llap{\large $p_z$}}
    \put(-1.,2.4){\llap{\large $\boxed{\epsilon_1}$}}
    \put(-1.,0.4){\llap{\large $\boxed{\epsilon_2}$}}
    
    \includegraphics[trim=0 180 0 120,clip,width=0.8\columnwidth]{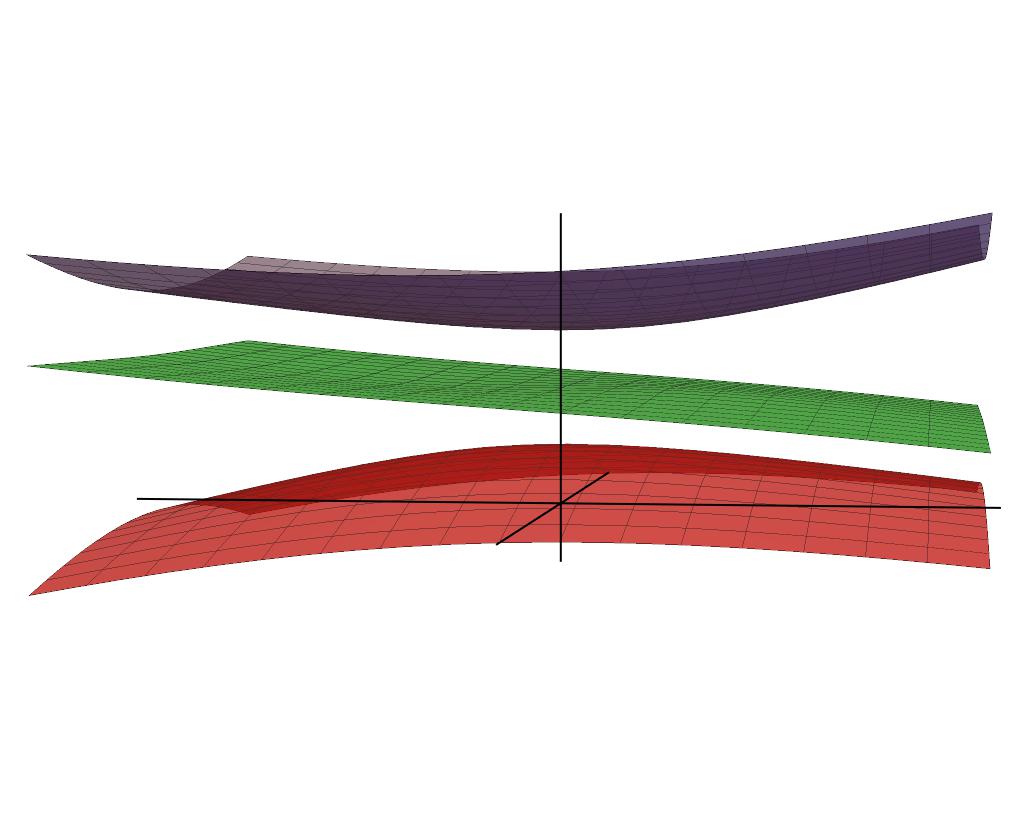}
    \setlength{\unitlength}{0.84cm}
    \put(0.5,1.0){\llap{\large $p_x$}}
    \put(-4.25,0.55){\llap{\large $p_y$}}
    \put(-3.5,3.75){\llap{\large $\epsilon_1$}}    
    
    \includegraphics[trim=0 180 0 120,clip,width=0.8\columnwidth]{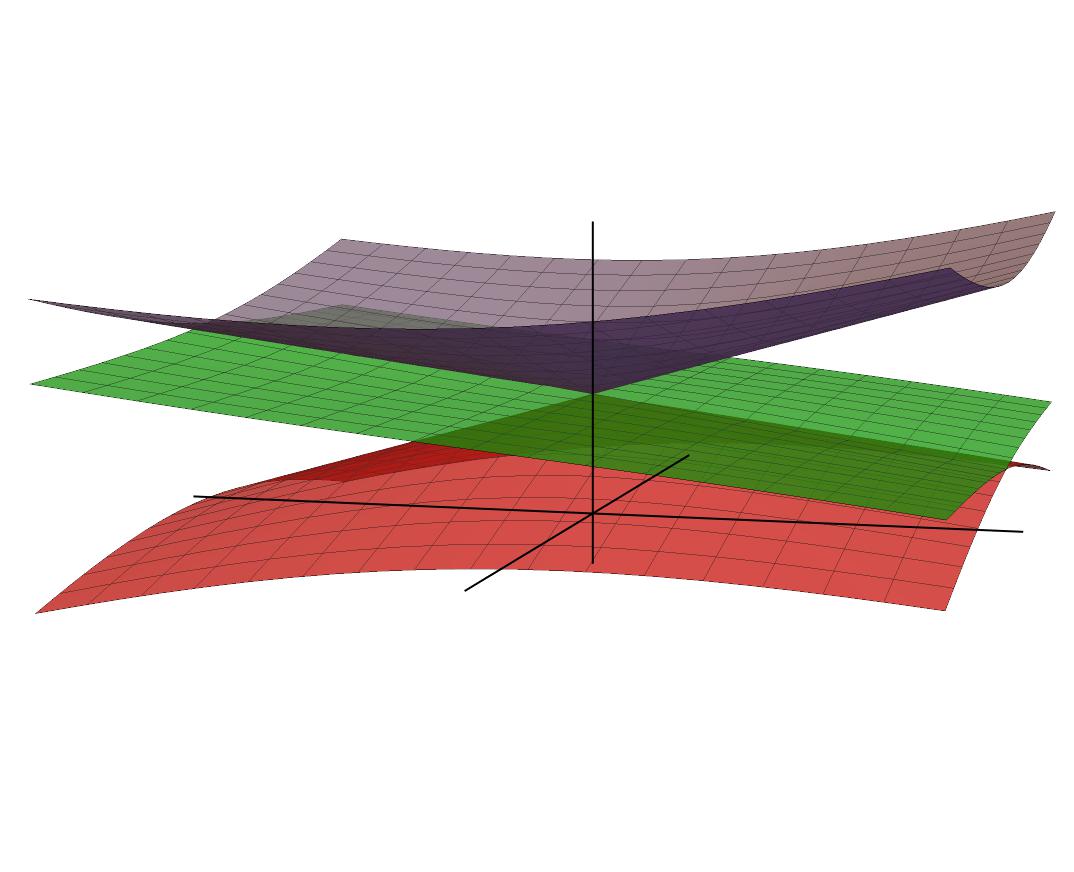}
    \setlength{\unitlength}{0.84cm}
    \put(0.2,1.1){\llap{\large $p_x$}}
    \put(-4.65,0.6){\llap{\large $p_y$}}
    \put(-3.5,3.75){\llap{\large $\epsilon_2$}}
    
    \caption{
    This figure shows
    cuts with the fine tuning such that the plane is parallel to
    the line degeneracy
    instead of generic cuts as in  Fig. \ref{fig:ham_lambda3z}. 
    There are only two types of cuts and the corresponding
    $2d$ band structure are shown in the following panels.}
    \label{fig:ham_lambda3z_perp_cut}
\end{figure}
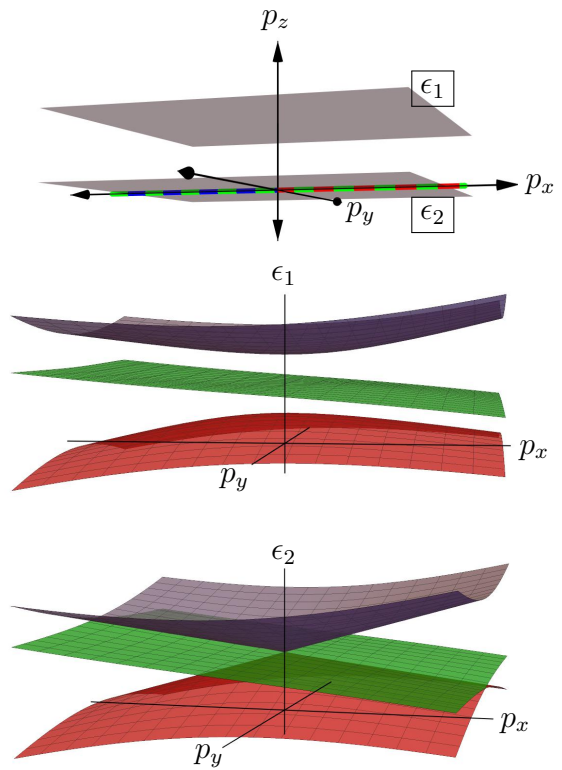

To write down the analyticity relations for the $H^3$ triple
point, we will again be motivated by how the family of $2d$
systems on cross-sections that include the triple point
interpolate among each other. There are two such examples
one shown in Fig. \ref{fig:ham_lambda3z} and another shown in
Fig. \ref{fig:ham_lambda3z_perp_cut}.
We see that certain cross-sections will resemble the
$2d$ Nexus system (as in Fig. \ref{fig:ham_lambda3z_perp_cut}),
while certain cross-sections will resemble a
$SU(2)$ spin-1 system (as in Fig. \ref{fig:ham_lambda3z}).

For the $2d$ Nexus-like cross-sections, the analyticity relations
are given by Eq. \ref{eq:2dnexus_analyticity_bc} (and
\ref{eq:2dnexus_analyticity_bp}). While for the $2d$ spin-1
cross-sections, they are 
\begin{subequations}
\begin{align}
v_{\text{top}}(\theta+\pi)=v_{\text{bottom}}(\theta) \label{eq:2dspin1_analyticity_bc}\\
v_{\text{middle}}(\theta+\pi)=v_{\text{middle}}(\theta)    
\label{eq:2dspin1_analyticity_bp}.
\end{align}
\label{eq:2dspin1_analyticity}
\end{subequations}
and clearly also the relation $v_i(\theta+2\pi)=v_i(\theta)$.
We note here that Eq. \ref{eq:2dspin1_analyticity_bp} captures the
spin-1 nature as opposed to a two-fold Dirac degeneracy and a third
standalone band.

To give a different example, we quickly look at the case of adding 
a diagonal $\Lambda^8$  ``mass" term
\begin{equation}
    H^8(\mathbf{p})=H(\mathbf{p})+p_z \Lambda^8 \label{eq:ham_lambda8z}
\end{equation}
For this case, there is line degeneracy
along $p_x$-axis as well as $p_z$-axis connected
to the triple point degeneracy. (One can easily see the
$p_z$-axis degeneracy coming from the eigenspectrum 
of $H^8(p_x=0,p_y=0,p_z)$.) This is illustrated in
the top panel of Fig. \ref{fig:H8_degenracy}. 
Generic cross-sections for $H^8(\mathbf{p})$ will contain 
two Dirac points either on the same pair of bands, or on 
different pairs of bands \emph{always involving} the middle band.
We can again define analyticity relations 
similar to Eqns. \ref{eq:2dnexus_analyticity_bc},
\ref{eq:2dnexus_analyticity_bp}, \ref{eq:2dspin1_analyticity}
for corresponding cross-sections containing the triple point. 

Finally, we end this section with the generalized domain
restatement for the $3d$ Nexus systems discussed above. It will
consist of \emph{three} copies of $p_x-p_y-p_z$ space which are 
joined appropriately at the line degeneracies (for both $H^3$
and $H^8$) and the triple point, with the above analyticity
relations giving us unambiguous rules to move in this generalized
domain. In the  next section -- where we build a \ADEDITOKAY{classification}{characterization}
scheme for Nexus triple points -- we will restrict ourselves to a
$d-1=2$ dimensional closed surface enclosing the triple point as
is done for the Weyl case. Again the analyticity relations will come to our
aid to govern how to move smoothly in this (generalized) $2d$
surface.


\section{Characterization}
\label{sec:classification}

In the previous sections, we established the rules to move
smoothly in our parameter space. Here, parameter space refers to
the generalized domain.
In this section, we will describe a (topological) \ADEDITOKAY{classification}{characterization}
scheme for different kinds of Nexus triple points by making use
of these rules. Given a Nexus system, the basic idea will be to 
consider an enclosing surface around the triple point in the 
generalized domain. As remarked
at the end of the previous section, the enclosing surface 
in the generalized domain consists of three copies of the surface
(e.g. spheres) joined at the points where the line degeneracies 
cross them. On this generalized enclosing surface, we will
categorize the various topologically distinct
ways in which one may analytically loop back to the start point. 
This is reminiscent of the concept of 
homology classes of 1-cycles \cite{Munkres_2013} in topological
classification of geometric objects. 
A very familiar example of this are the non-trivial loops that 
one draws on a torus that can not be shrunk to a point, whereas 
on a sphere there are no such loops. 
Importantly, the analyticity relations discussed 
before allows us to focus only on the enclosing surface 
to capture the topological data of the wavefunction geometry 
without the full knowledge of the wavefunctions themselves.

\begin{figure}[h]
\centering
\includegraphics[trim=250 300 100 100,clip,width=0.7\columnwidth]{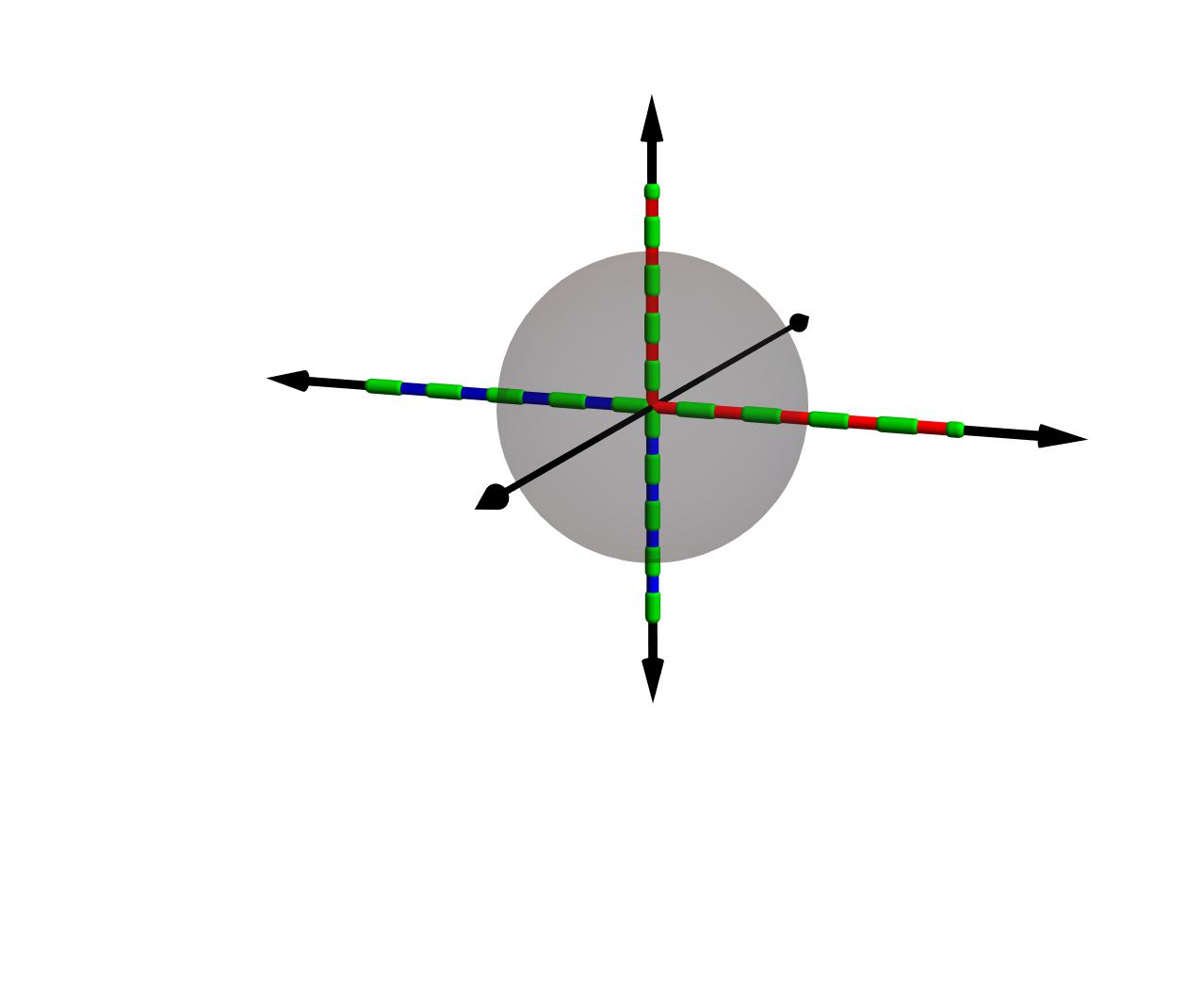}
\setlength{\unitlength}{0.74cm}
\put(0.55,2.65){\llap{\large $p_x$}}
\put(-2.1,3.9){\llap{\large $p_y$}}
\put(-4.,6.25){\llap{\large $p_z$}}

\includegraphics[trim=0 150 0 180,clip,width=0.8\columnwidth]{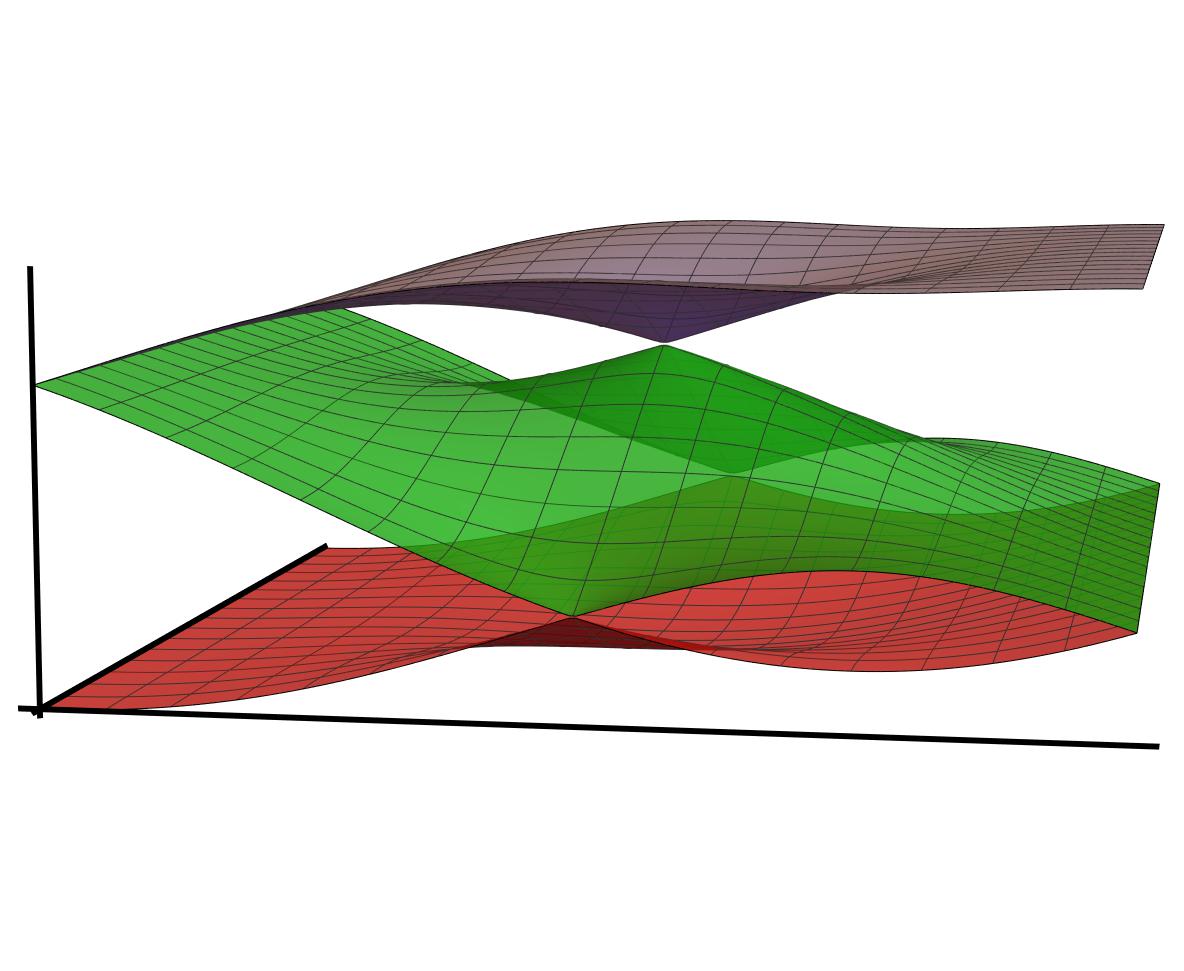}
\setlength{\unitlength}{0.85cm}
\put(-4.,0.){\llap{\large $\theta$}}
\put(-7.,1.5){\llap{\large $\phi$}}
\put(-7.85,4.0){\llap{\large $\epsilon$}}
\vspace*{0.5cm}
\includegraphics[trim=50 70 50 80,clip,width=0.265\columnwidth]{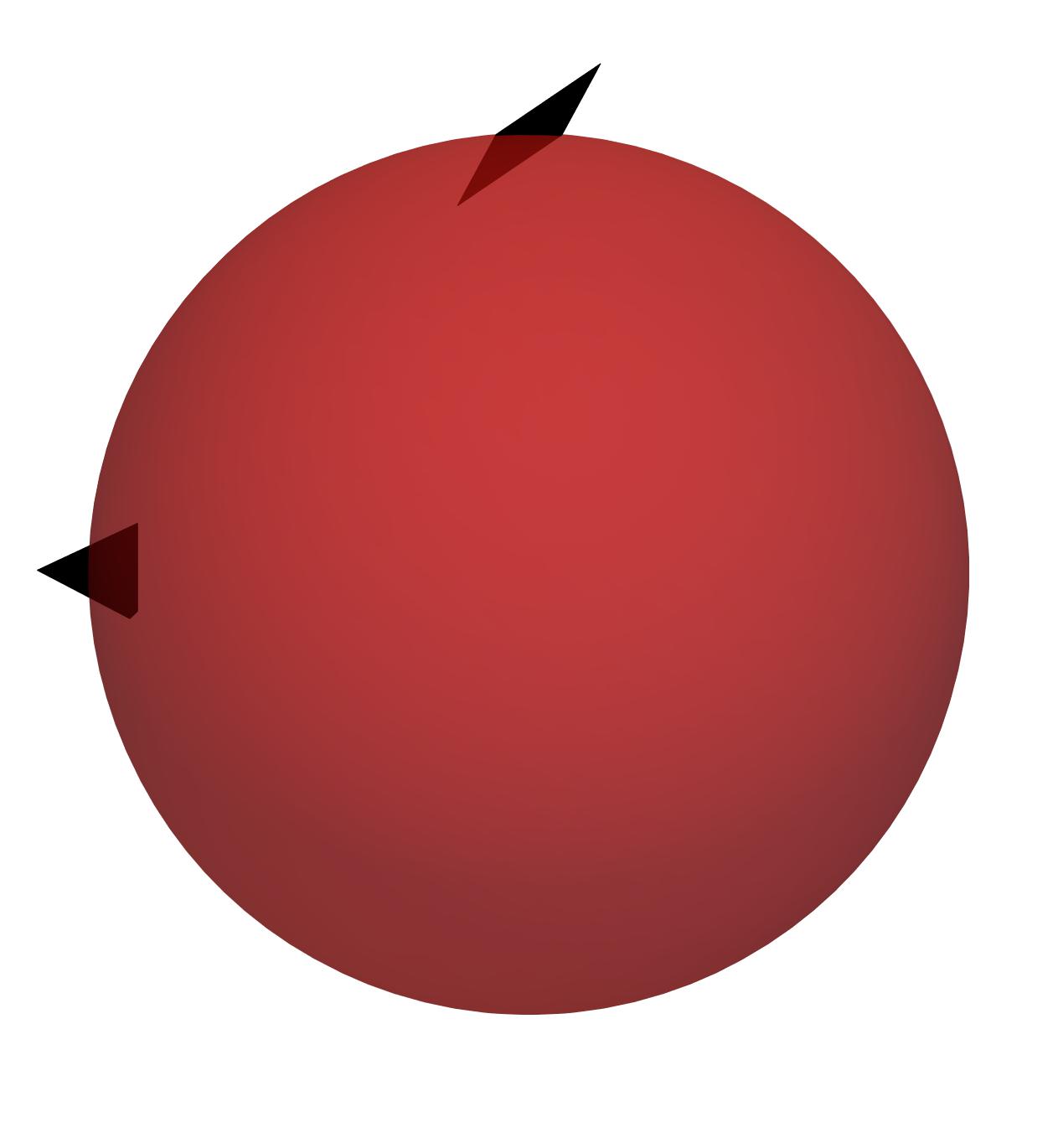}
\includegraphics[trim=50 70 50 80,clip,width=0.265\columnwidth]{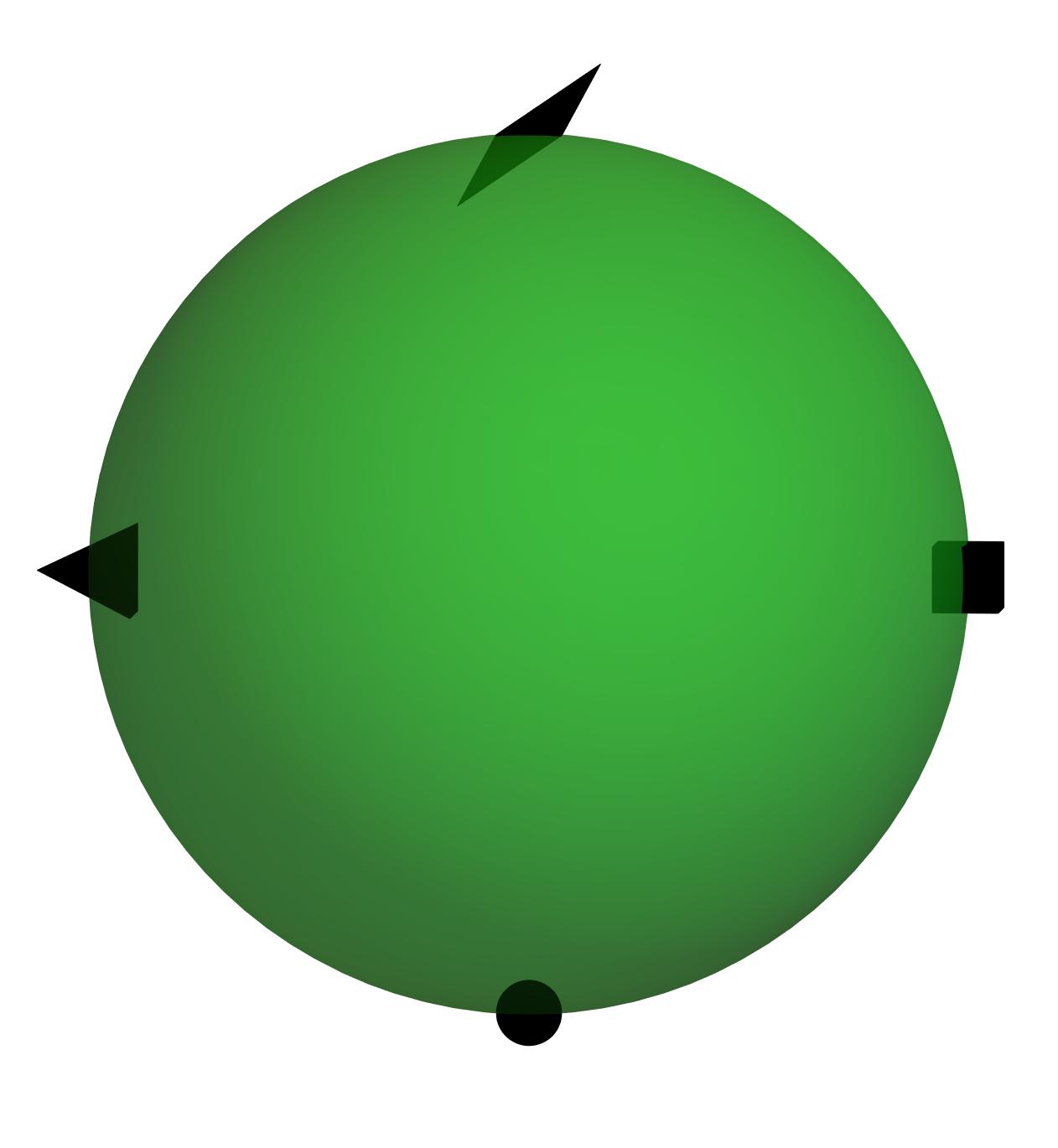}
\includegraphics[trim=50 70 50 80,clip,width=0.265\columnwidth]{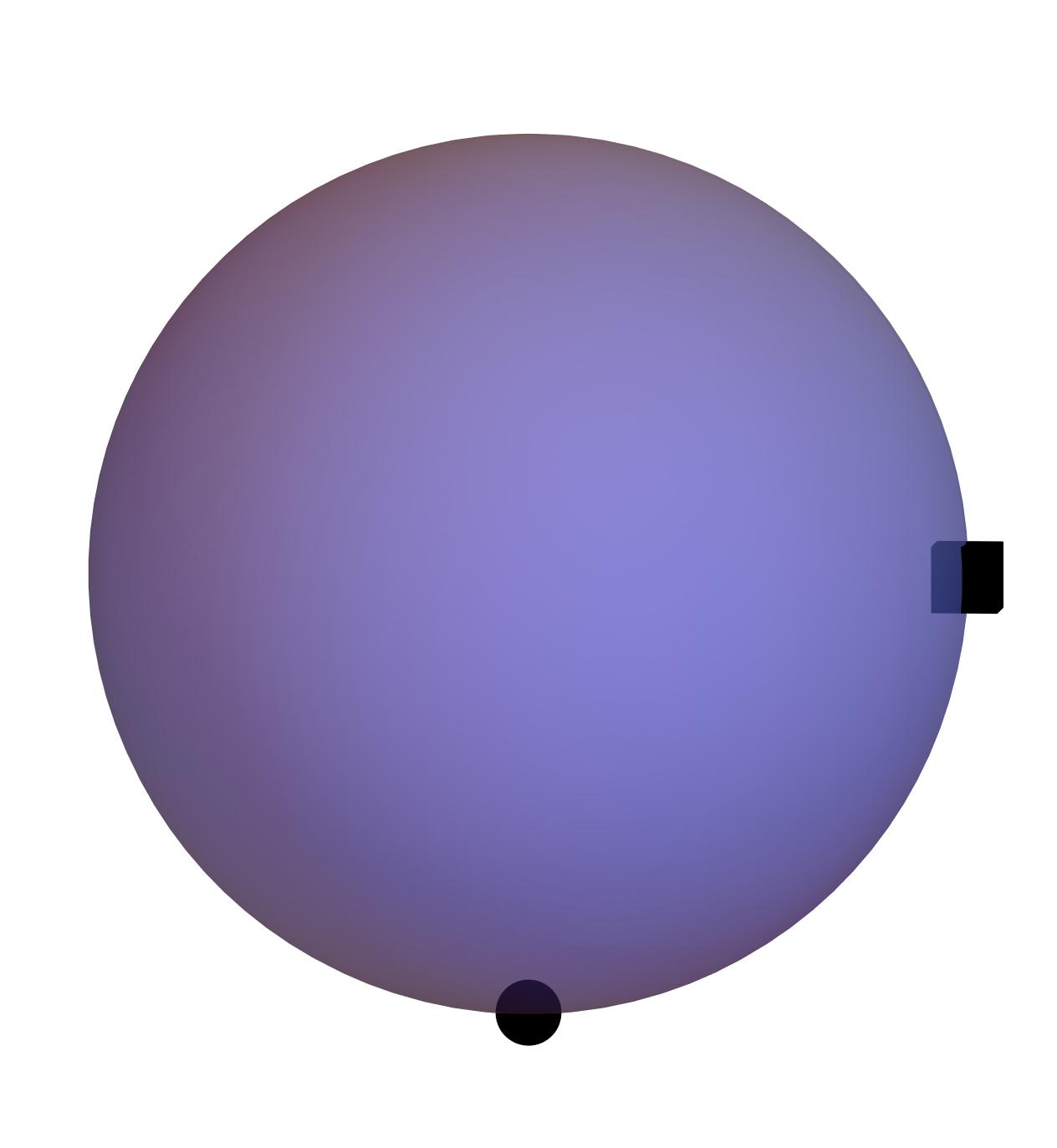}

\caption{
The first panel shows the line degeneracies for
$H^8$ using similar convention as Fig. \ref{fig:ham_lambda3z}. 
The enclosing surface on the original domain
is also shown as the grey sphere. The second panel  
is the plot of the energy spectrum on the
enclosing surface parametrized by $\theta,\phi$.
The third panel
shows the enclosing surface in the generalized domain 
which consists of
three copies of the original enclosing surface
connected to each other at the intersection points with
the underlying line degeneracies.
There are four different shaped points on these spheres 
representing the four connecting points.}
\label{fig:h8_enclosing_surface}
\label{fig:H8_degenracy}
\end{figure}

\begin{figure}
\centering
\boxed{\includegraphics[trim=50 70 50 80,clip,width=0.26\columnwidth]{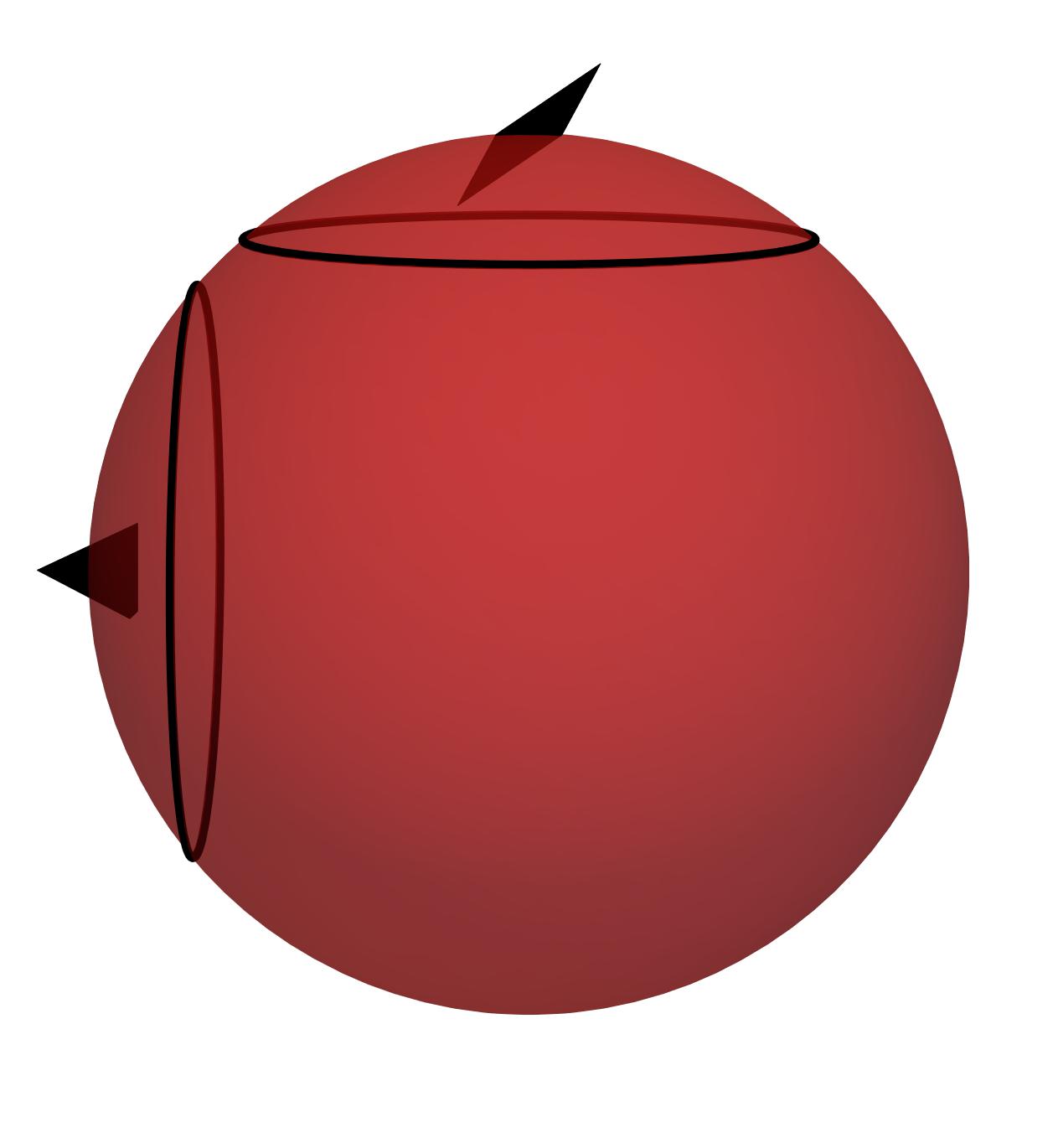}
\includegraphics[trim=50 70 50 80,clip,width=0.26\columnwidth]{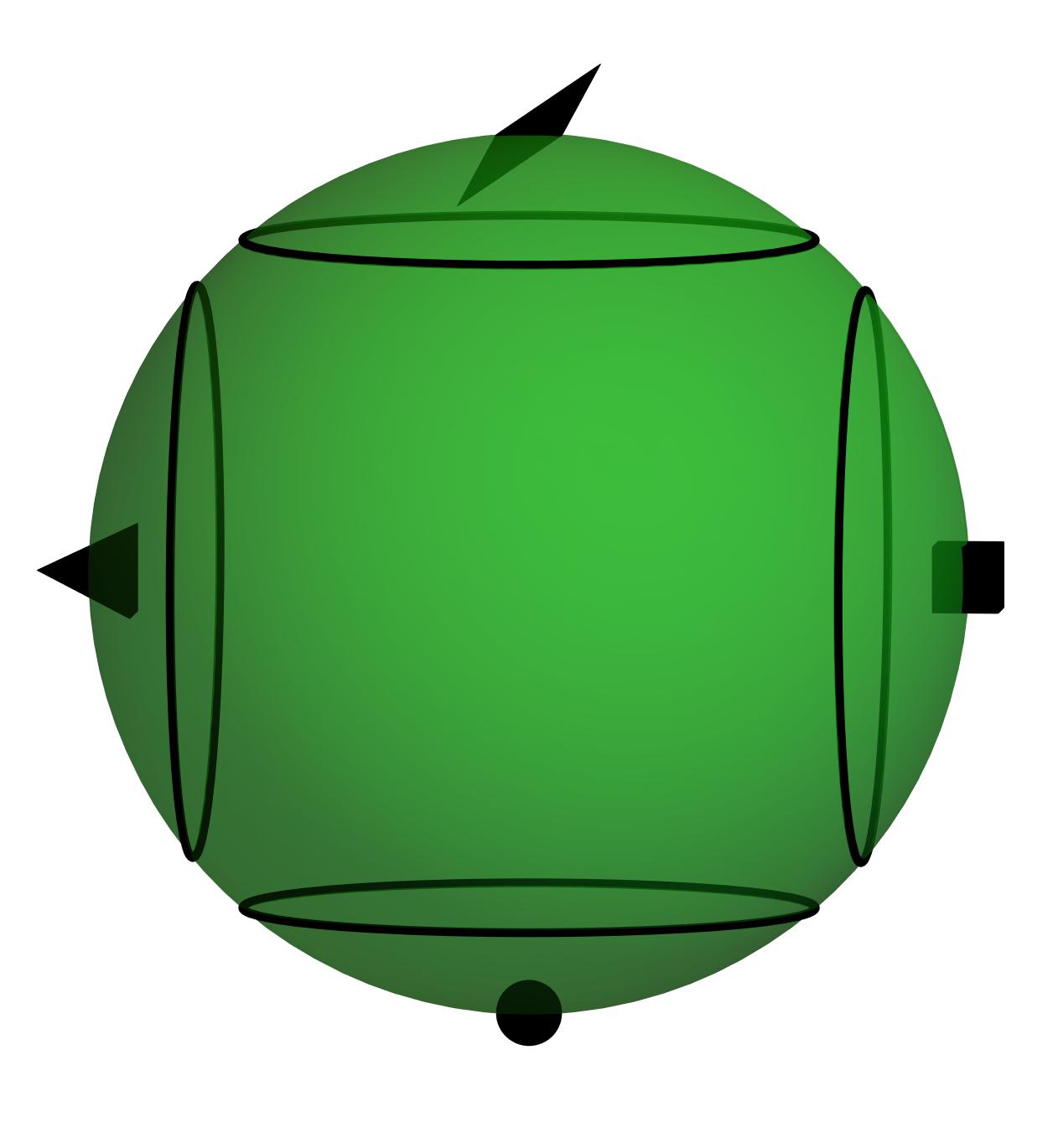}
\includegraphics[trim=50 70 50 80,clip,width=0.26\columnwidth]{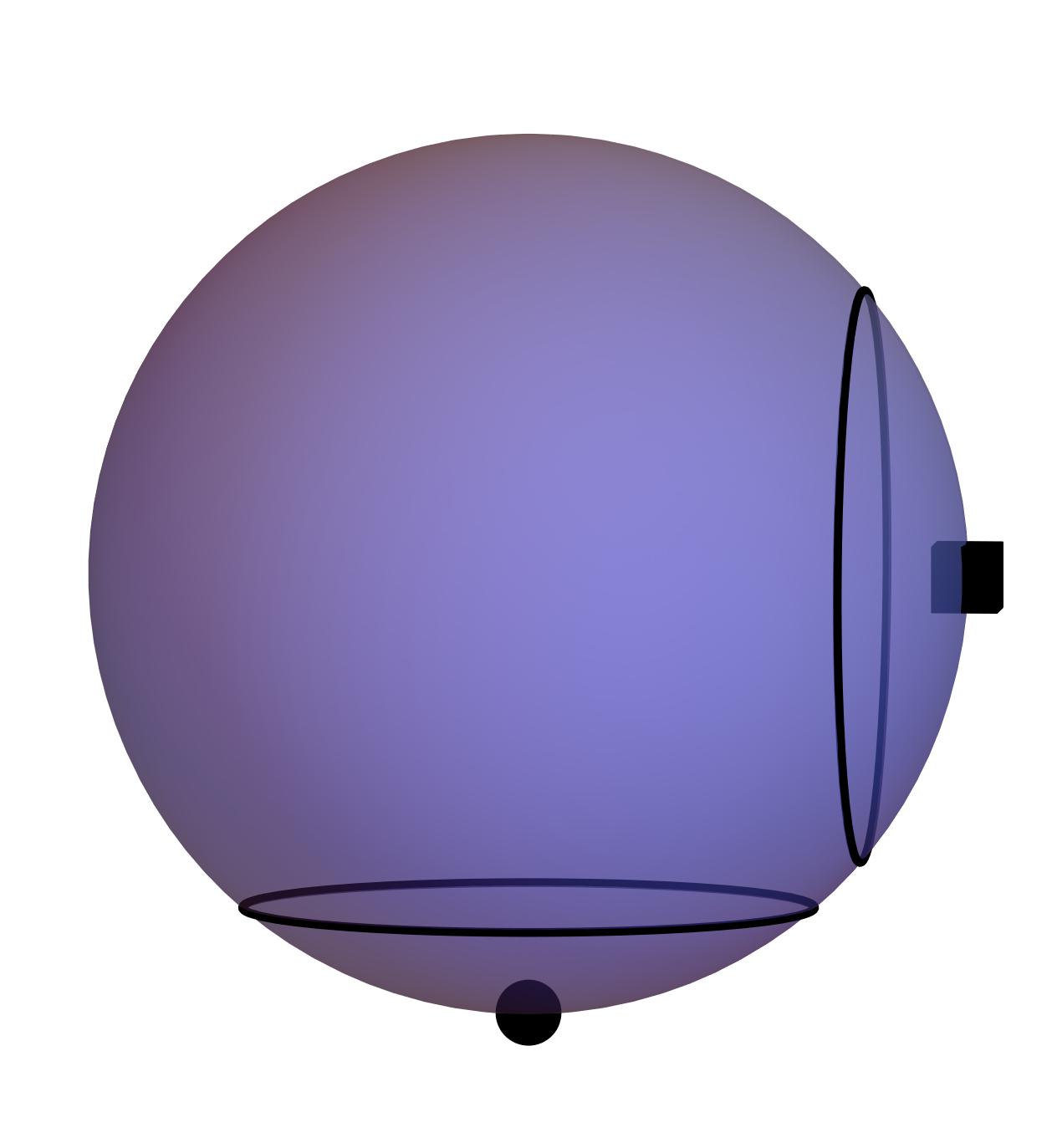}}
\boxed{\includegraphics[trim=50 70 50 80,clip,width=0.26\columnwidth]{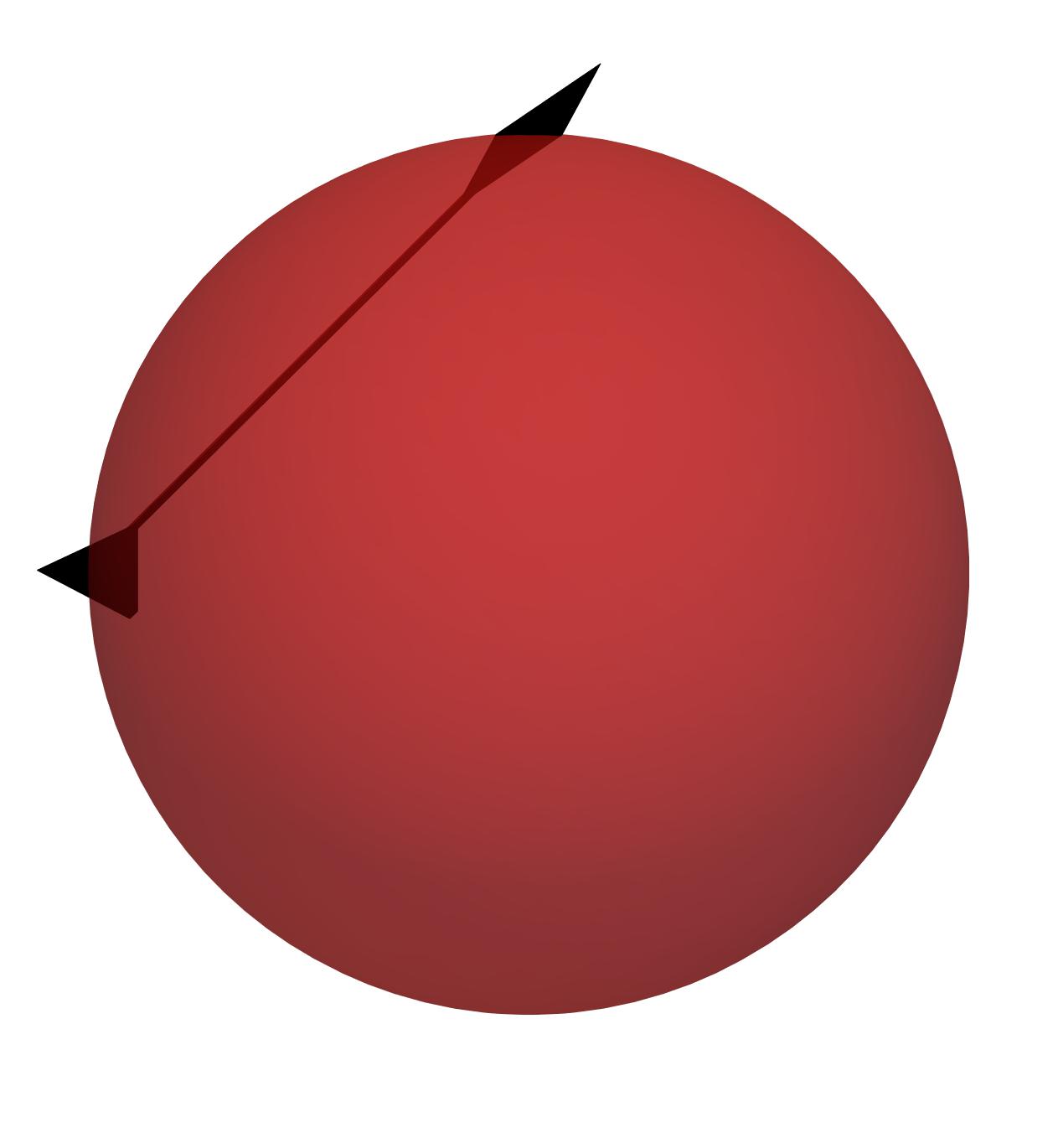}
\includegraphics[trim=50 70 50 80,clip,width=0.26\columnwidth]{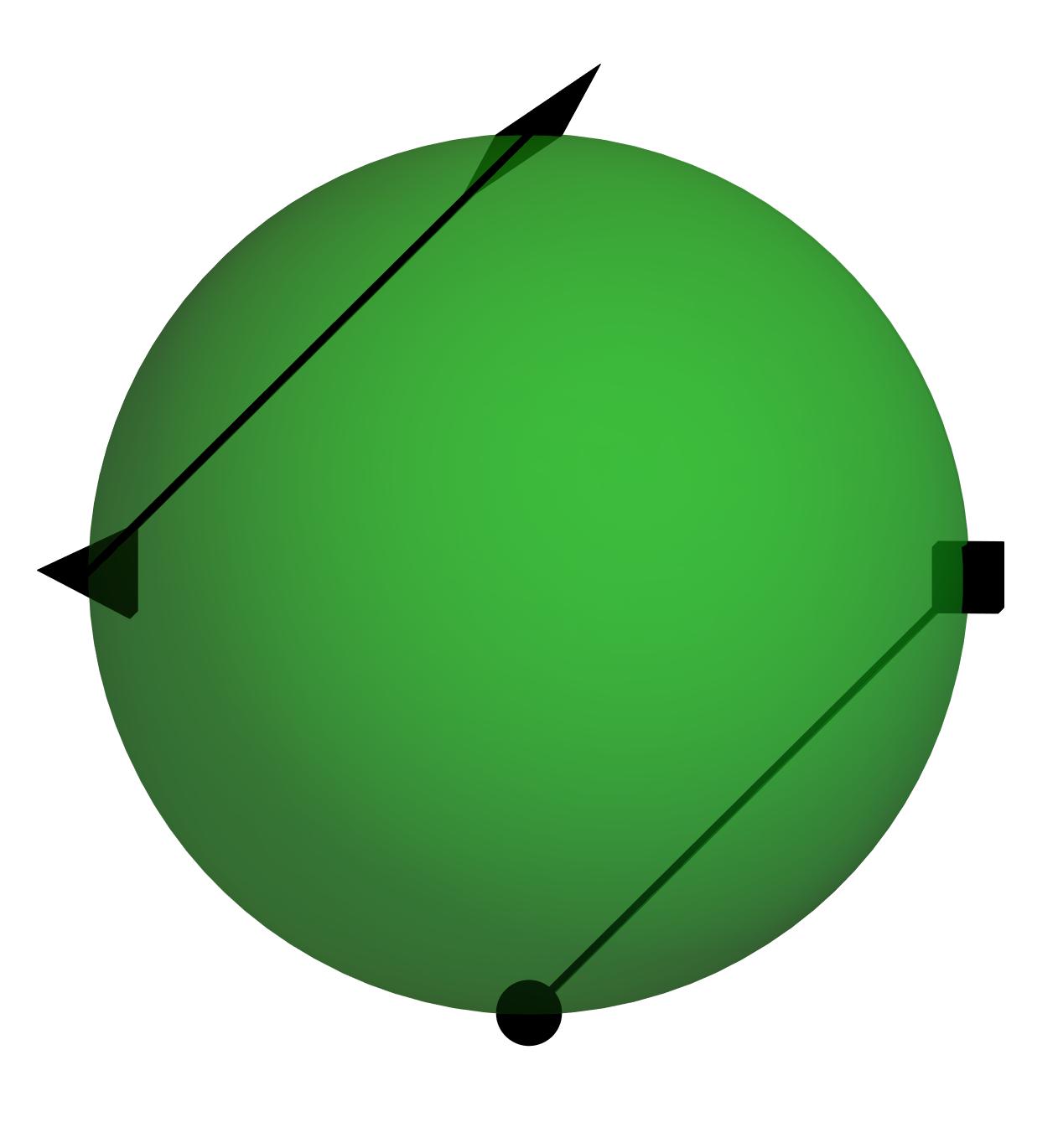}
\includegraphics[trim=50 70 50 80,clip,width=0.26\columnwidth]{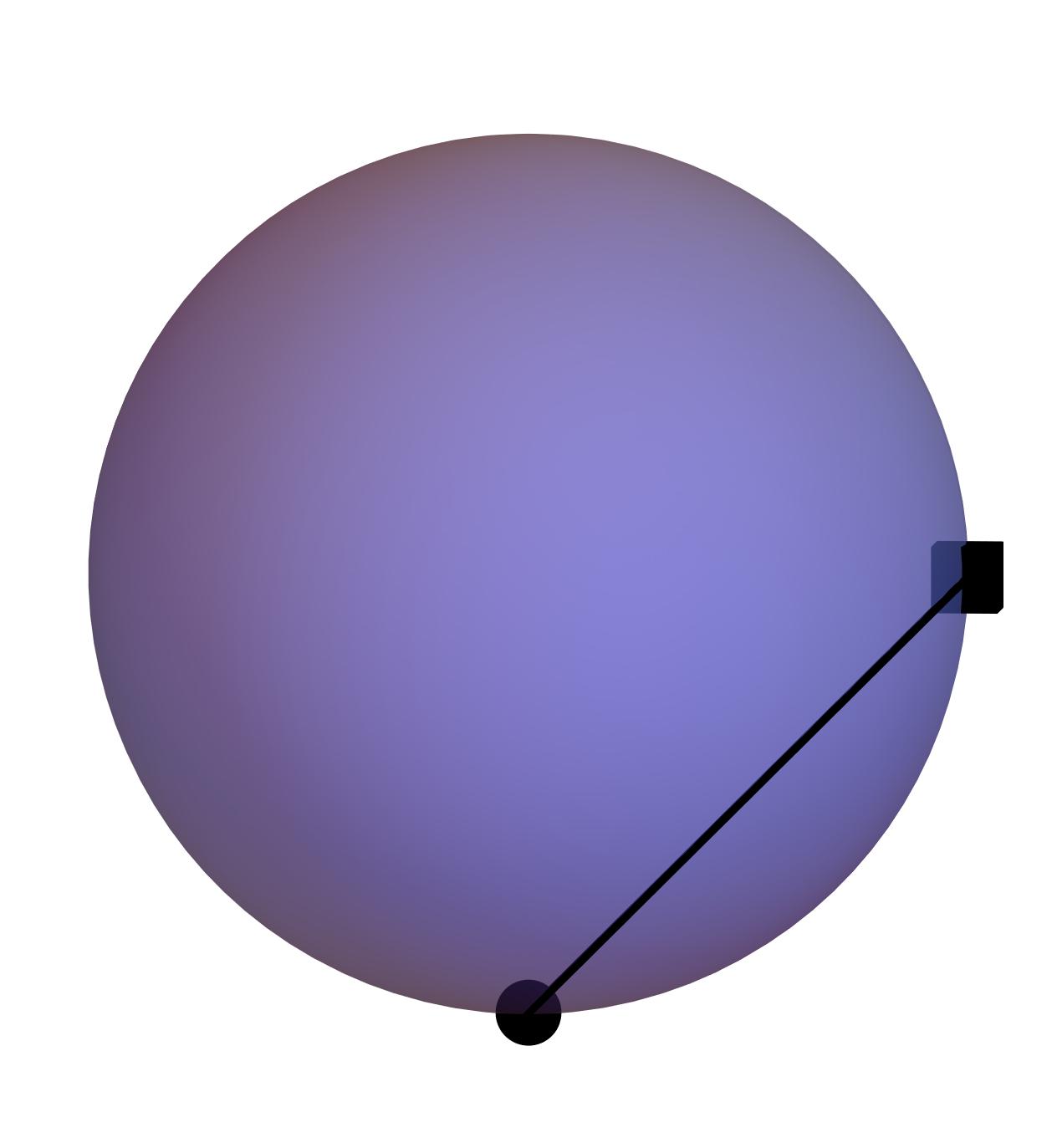}}
\boxed{\includegraphics[trim=50 70 50 80,clip,width=0.26\columnwidth]{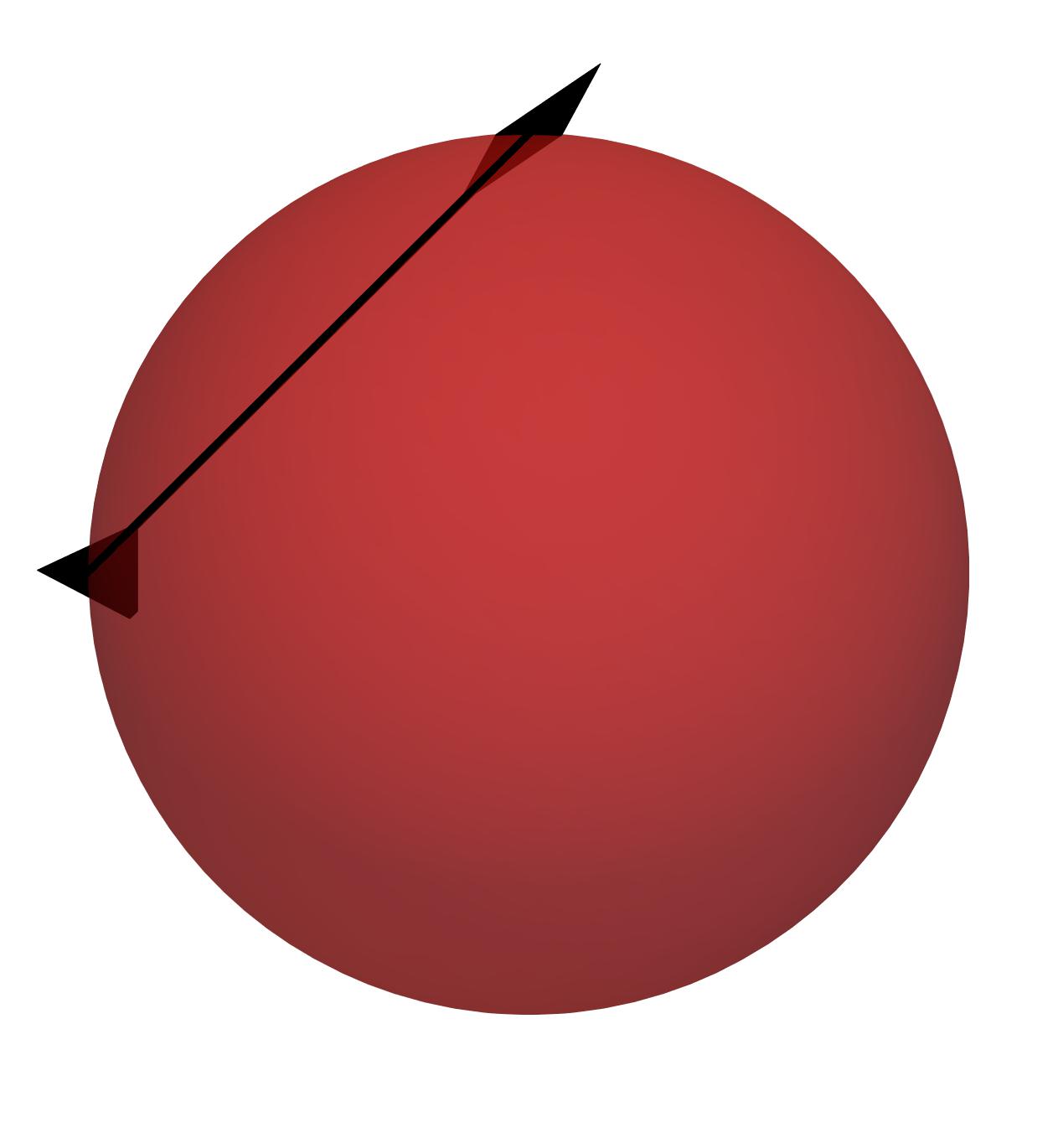}
\includegraphics[trim=50 70 50 80,clip,width=0.26\columnwidth]{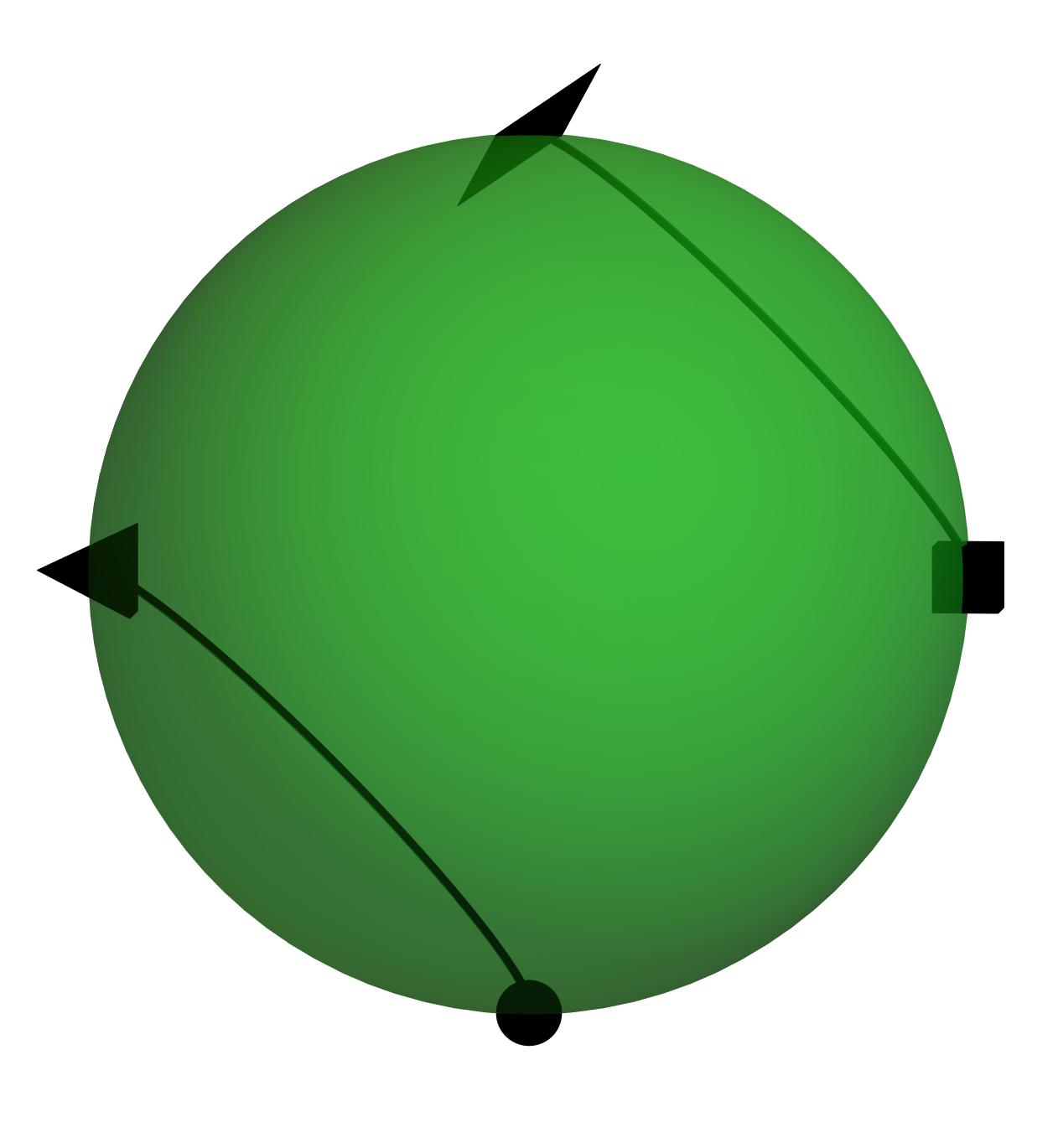}
\includegraphics[trim=50 70 50 80,clip,width=0.26\columnwidth]{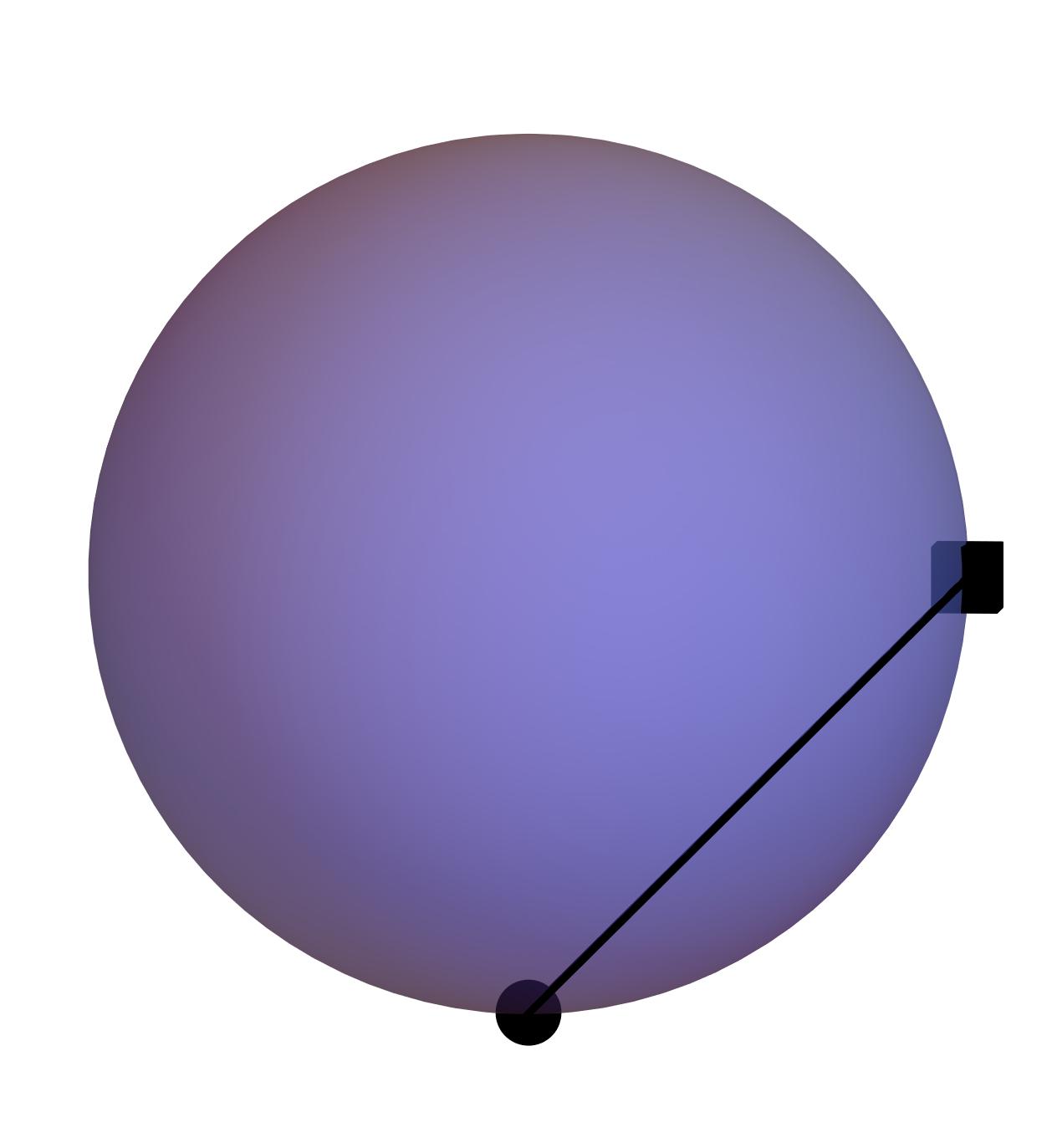}}
\boxed{\includegraphics[trim=50 70 50 80,clip,width=0.26\columnwidth]{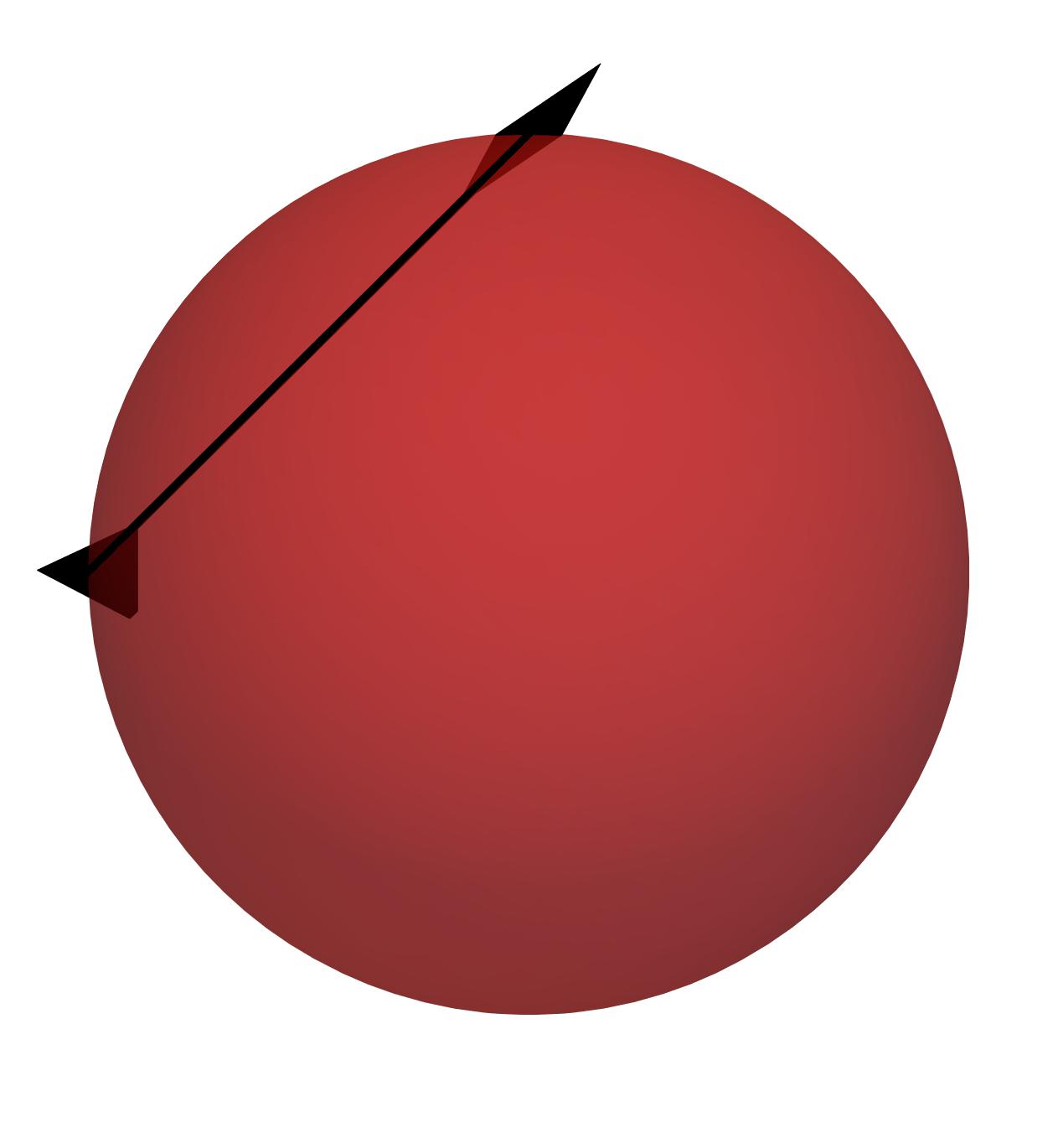}
\includegraphics[trim=50 70 50 80,clip,width=0.26\columnwidth]{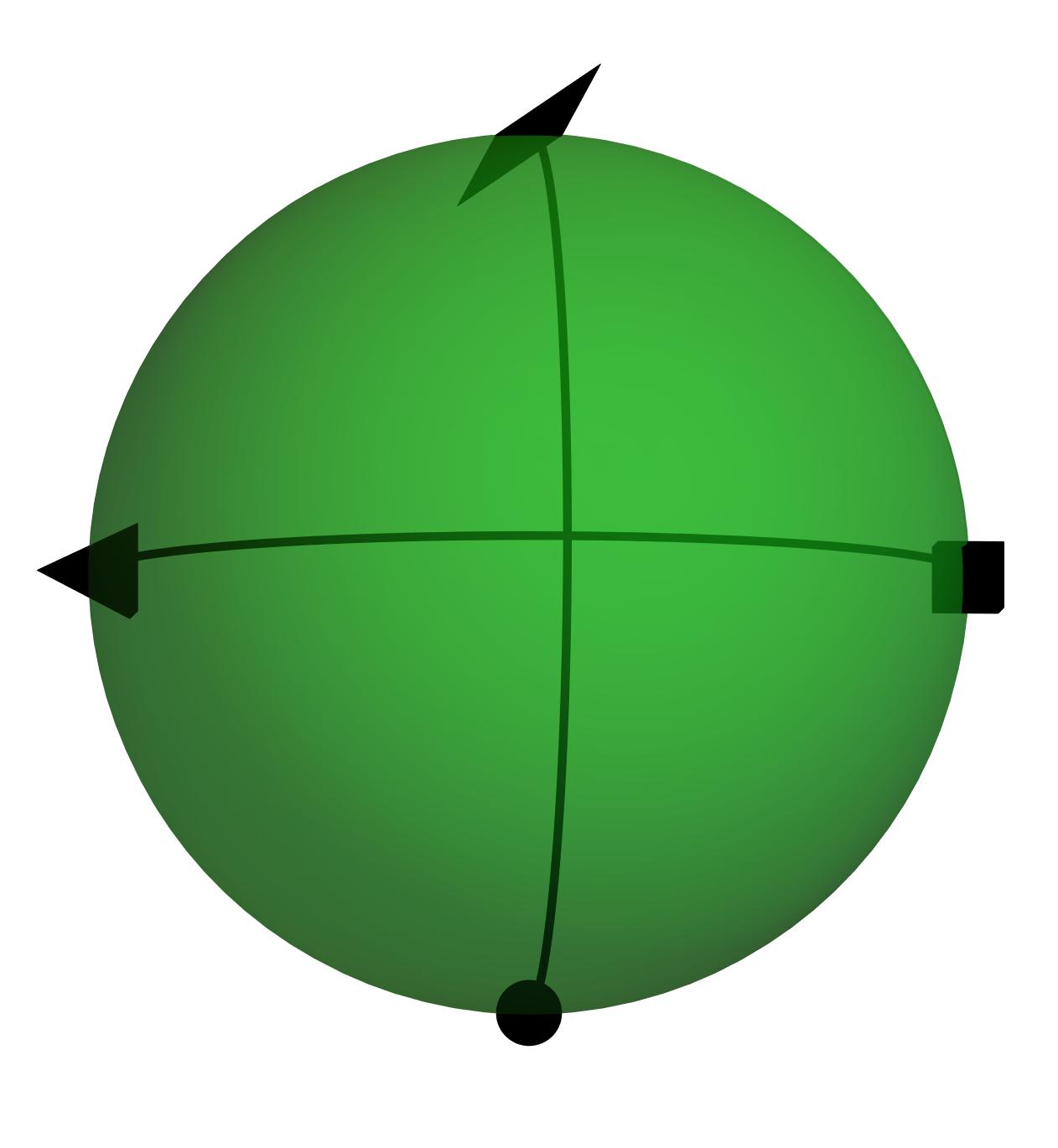}
\includegraphics[trim=50 70 50 80,clip,width=0.26\columnwidth]{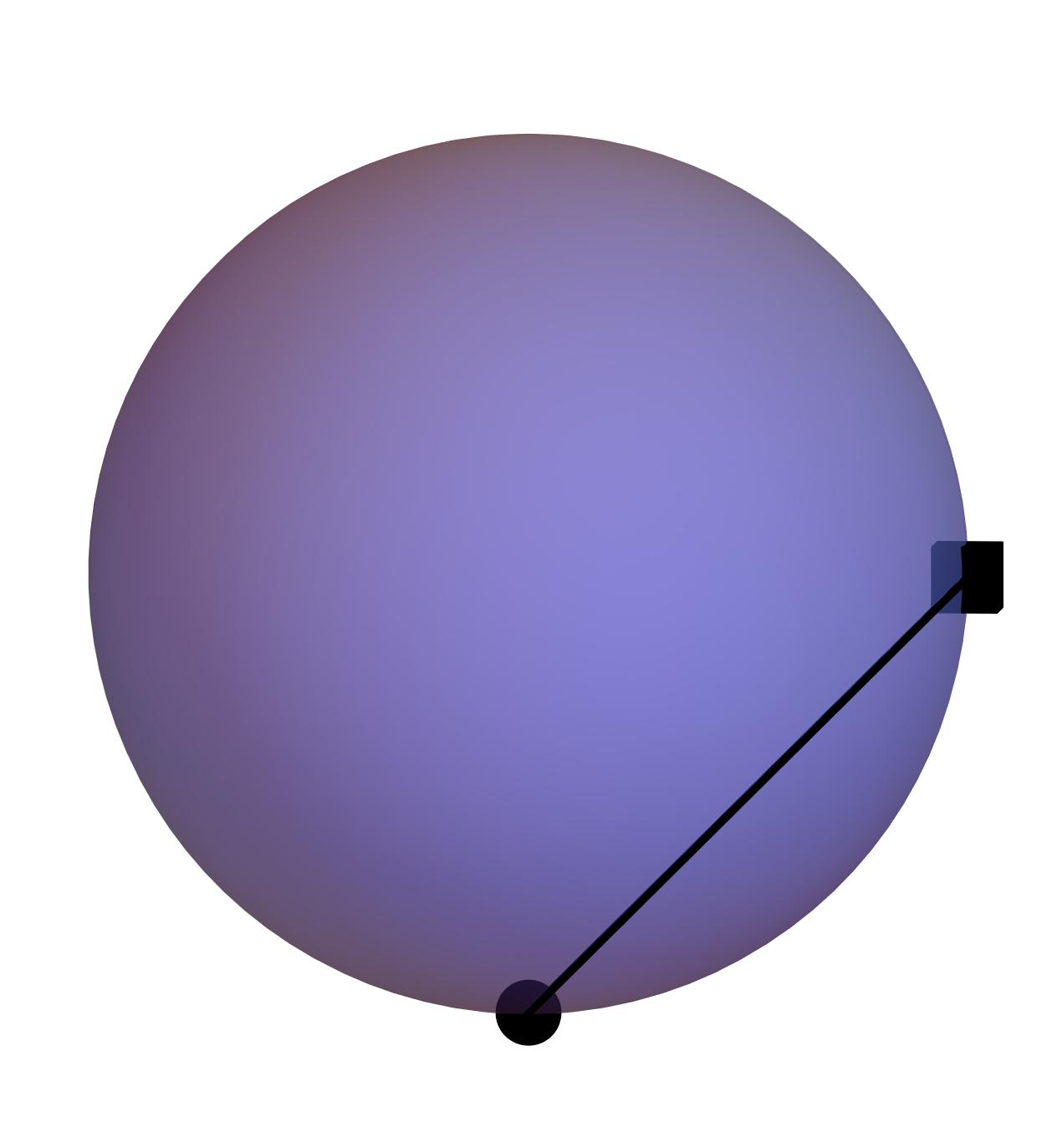}}
\caption{This figure shows the homological
characterization of non-contractible loops for the case of 
 $H^8$ triple point.
The first panel shows the loops 
which do not touch any connecting point. The second panel 
has all the loops which touch exactly two connecting points. 
Loops touching only one connecting point is not possible because
of Fig. \ref{fig:dirac_move_sketch}.
The third and the fourth panel shows two
different kinds of loops which touch all the connecting points.}
\label{fig:homology_h8}
\end{figure}

Let's start with $H^8$ in this case. The enclosing surface for 
this is shown in Fig. \ref{fig:h8_enclosing_surface}. Let us 
imagine drawing topologically distinct loops on this.
Clearly there exist (trivial) loops that can be shrunk to a point
(not shown in the figures). $H^8$ also hosts
non-trivial loops which are shown in
Fig. \ref{fig:homology_h8}. We see there are two kinds of loops:
\begin{enumerate}
\item those that stay on the same sphere.
The drawing of such loops relies on
the index-preserving kind of analytic relations.
\item those that straddle different spheres.
The drawing of such loops relies on
the index-connecting kind of analytic relations.
\end{enumerate}
Close to the connecting point on the $2d$ enclosing surface,
we can imagine a small flat coordinate patch giving us our local 
coordinate system in which we may use Eq.
\ref{eq:dirac_analyticity_bc}. Therefore, in the drawing of the
loop through the connecting point, we \emph{have to use} the step
illustrated in Fig \ref{fig:dirac_move_sketch}. 
A corollary is that there can not be a non-trivial loop
on a single sphere that touches the Dirac-like connecting point.

\begin{figure}[h]
\centering
\includegraphics[trim=260 310 100 100,clip,width=0.7\columnwidth]{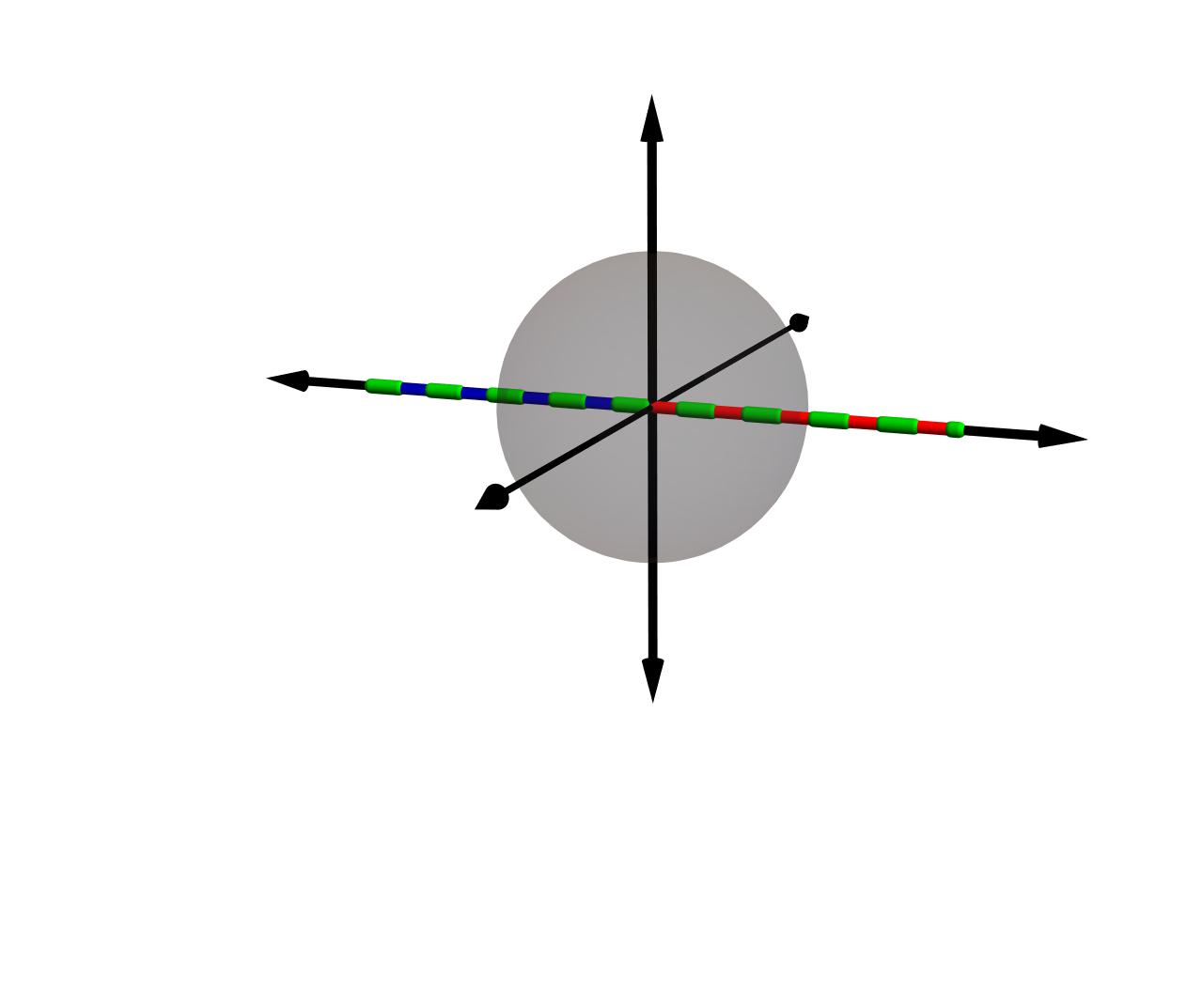}
\setlength{\unitlength}{0.74cm}
\put(0.55,2.55){\llap{\large $p_x$}}
\put(-2.1,4.){\llap{\large $p_y$}}
\put(-4.,6.2){\llap{\large $p_z$}}
\vspace*{0.5cm}

\includegraphics[trim=50 100 50 100,clip,width=0.265\columnwidth]{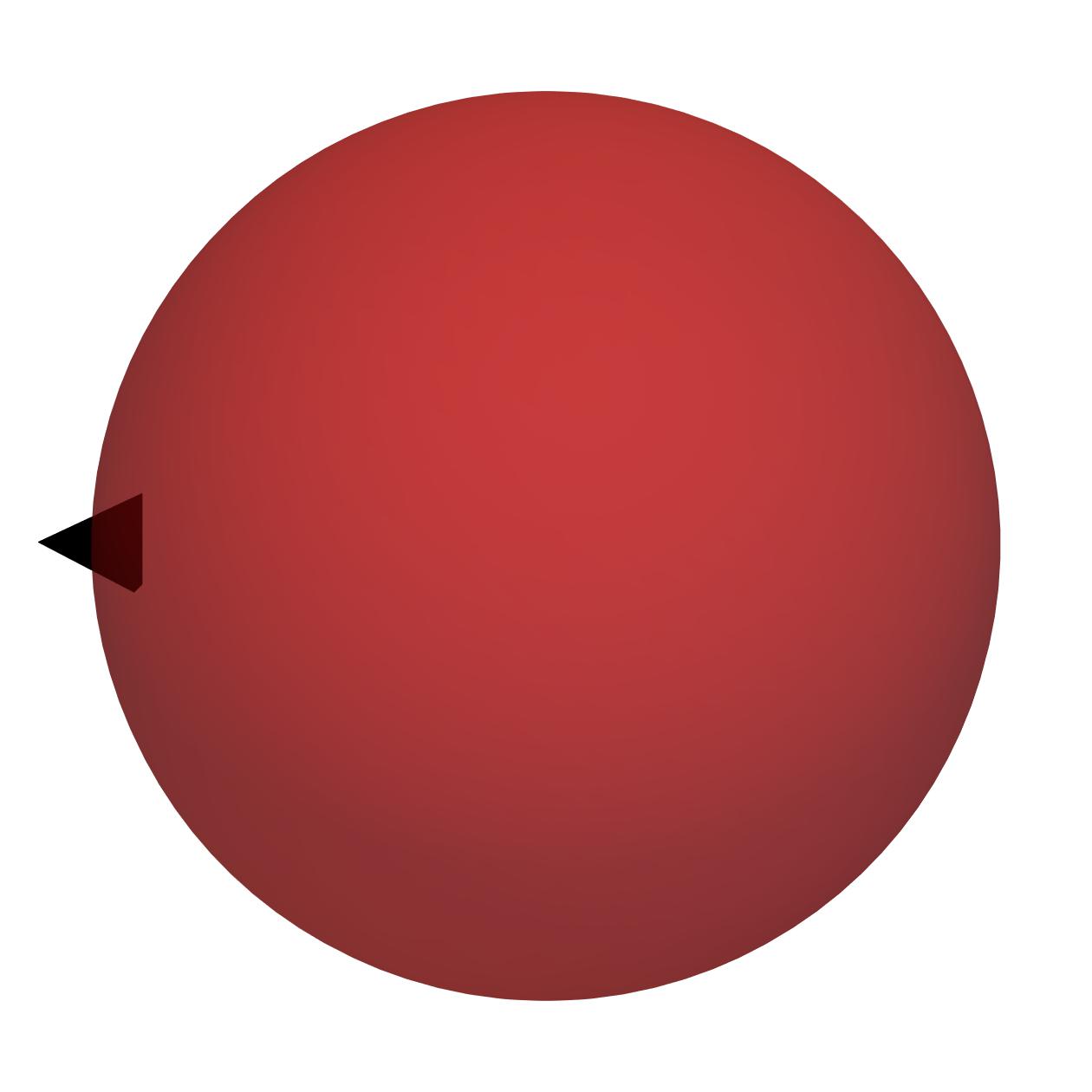}
\includegraphics[trim=50 100 50 100,clip,width=0.265\columnwidth]{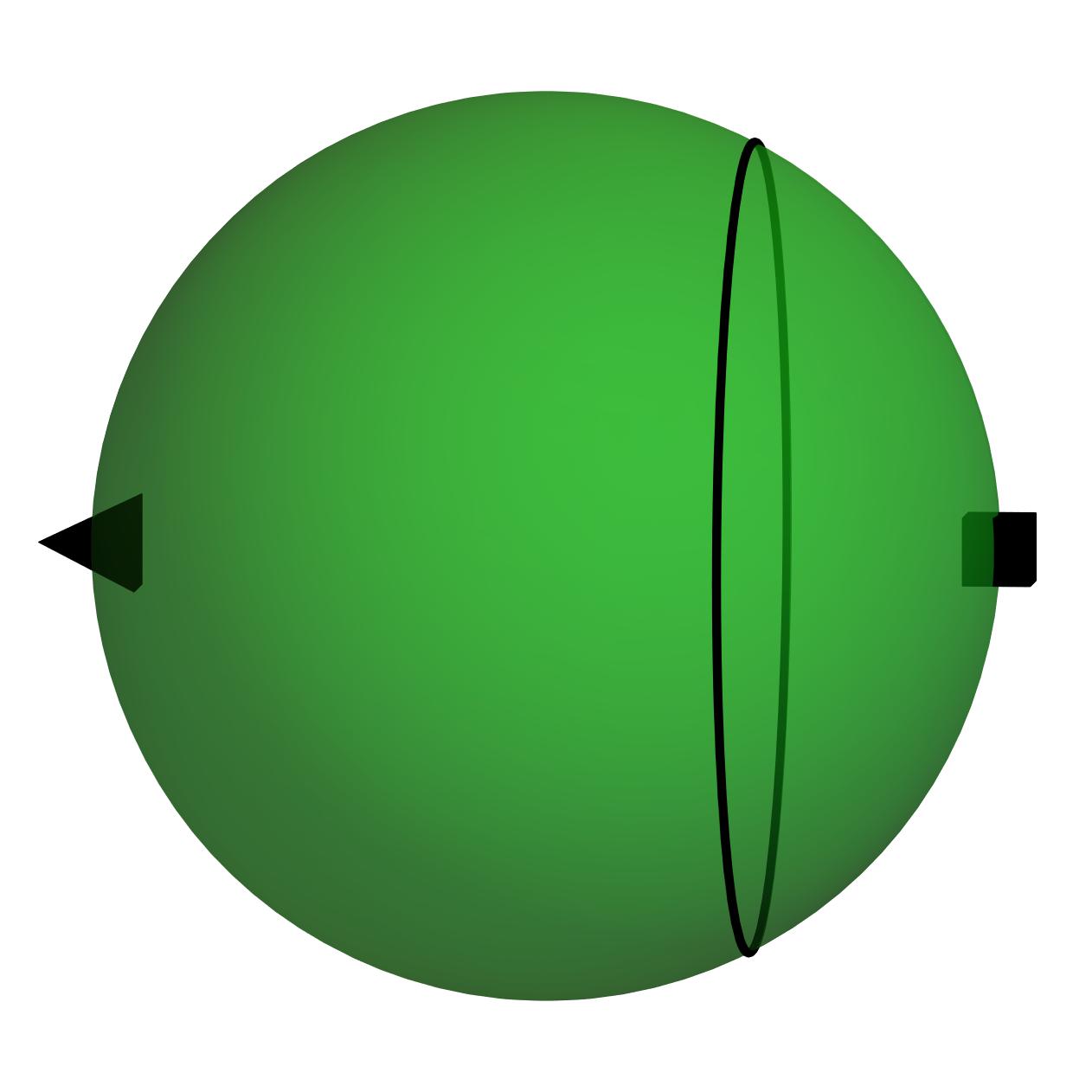}
\includegraphics[trim=50 100 50 100,clip,width=0.265\columnwidth]{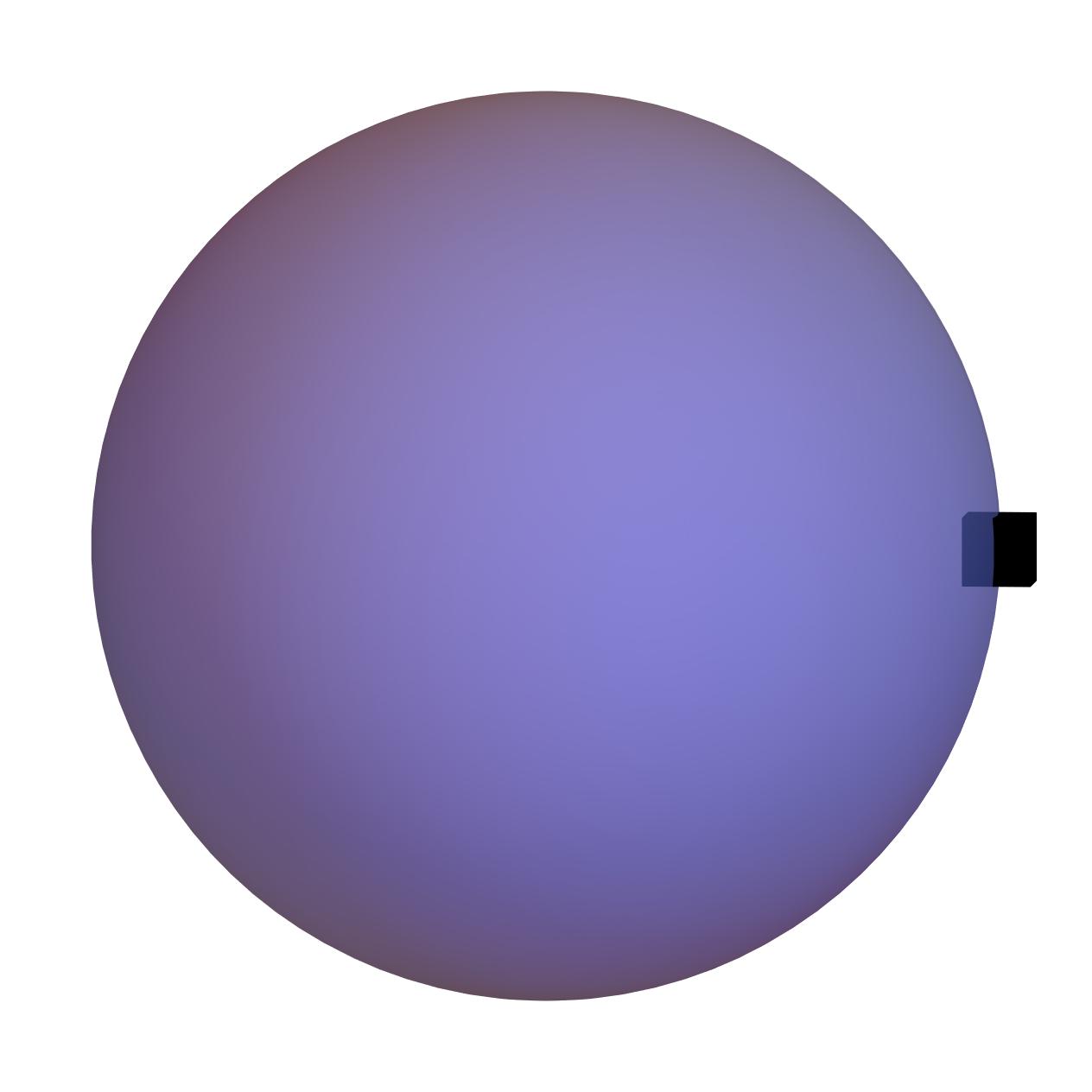}
\caption{The top panel shows the enclosing surface
in the original domain for $H^3$. 
In the generalized domain, the corresponding enclosing surface
consists of three connected spheres with associate connecting
points as shown in the second panel.
There is only one non-contractible loop one that
can be drawn in the middle sphere.
Any loop on the left and right spheres can be
contracted to a point. On the non-contractible loop,
one can calculate the Berry phase which will turn out be $\pm \pi$.}
\label{fig:H3_homology}
\end{figure}

\begin{figure}[h]
\centering
\boxed{\includegraphics[trim=50 70 50 80,clip,width=0.265\columnwidth]{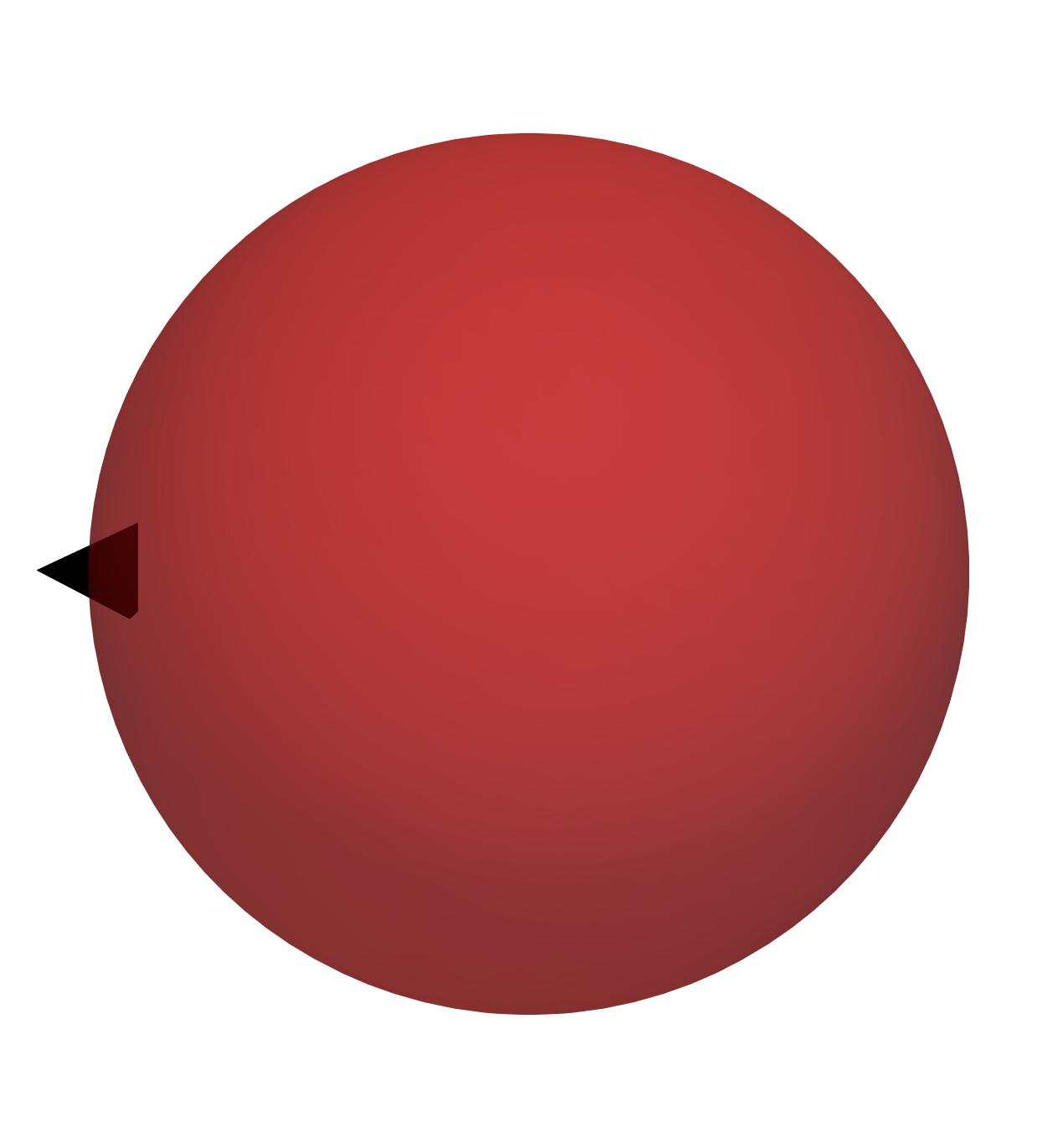}
\includegraphics[trim=50 70 50 80,clip,width=0.265\columnwidth]{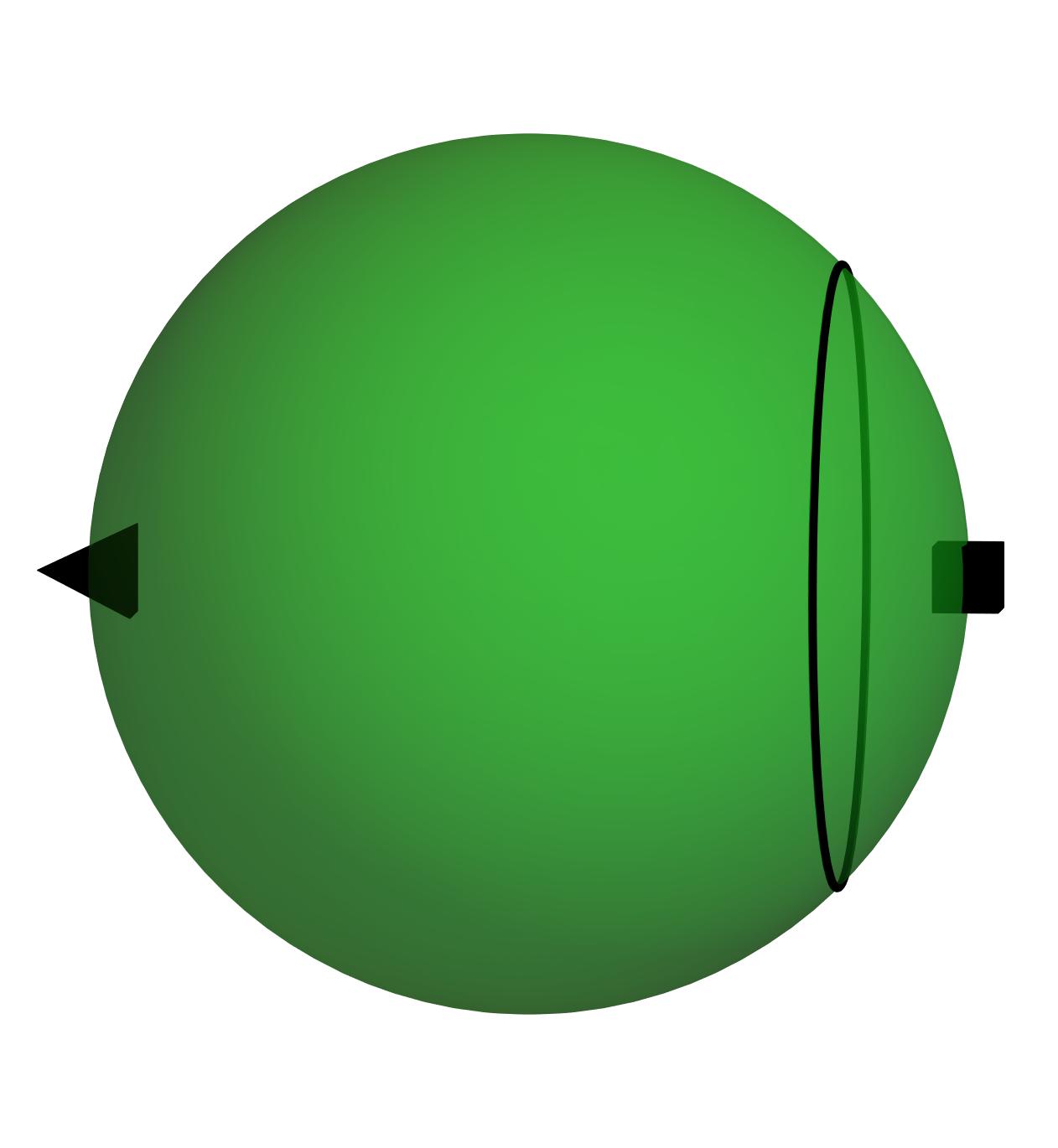}
\includegraphics[trim=50 70 50 80,clip,width=0.265\columnwidth]{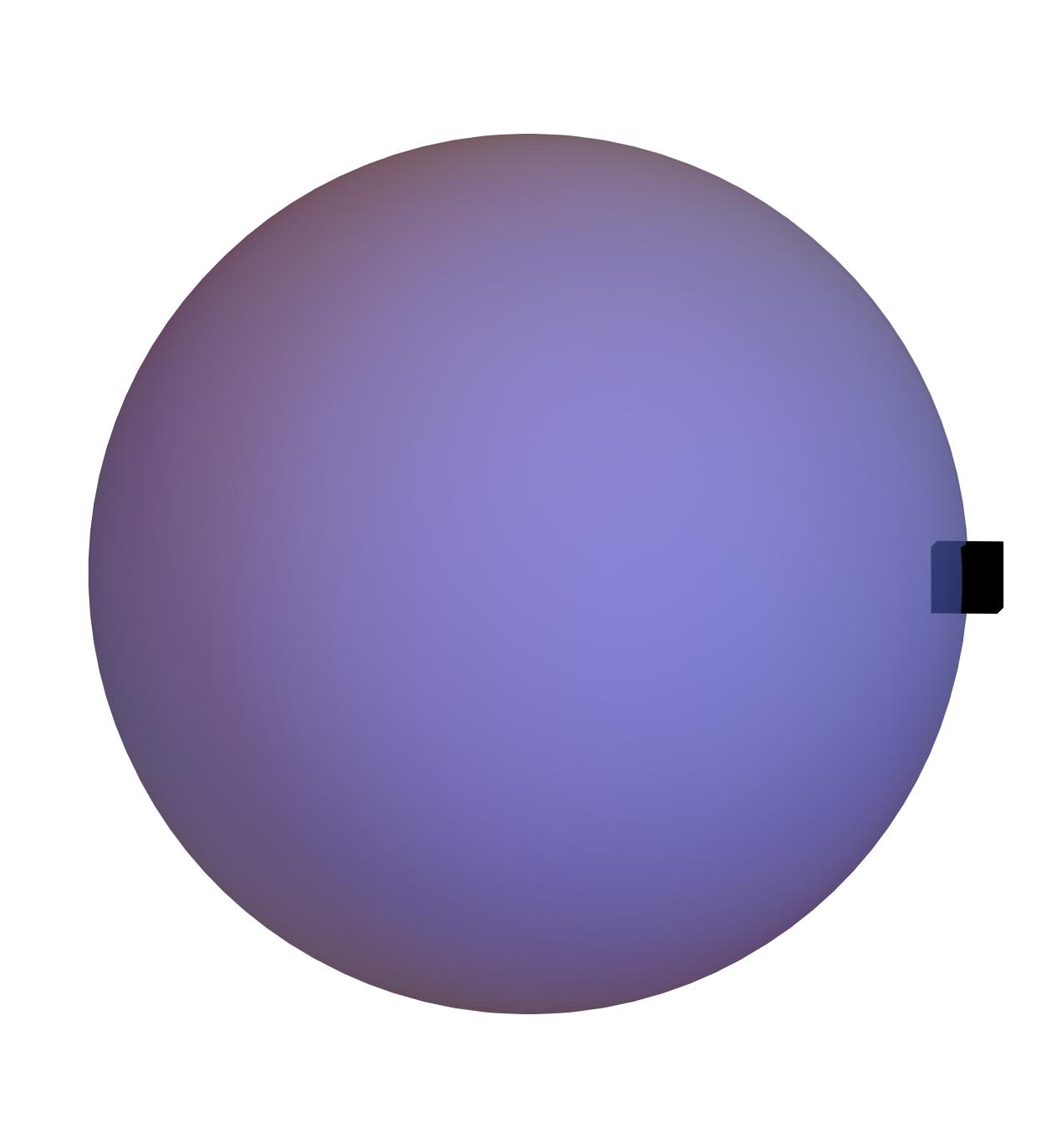}}
\boxed{\includegraphics[trim=50 70 50 80,clip,width=0.265\columnwidth]{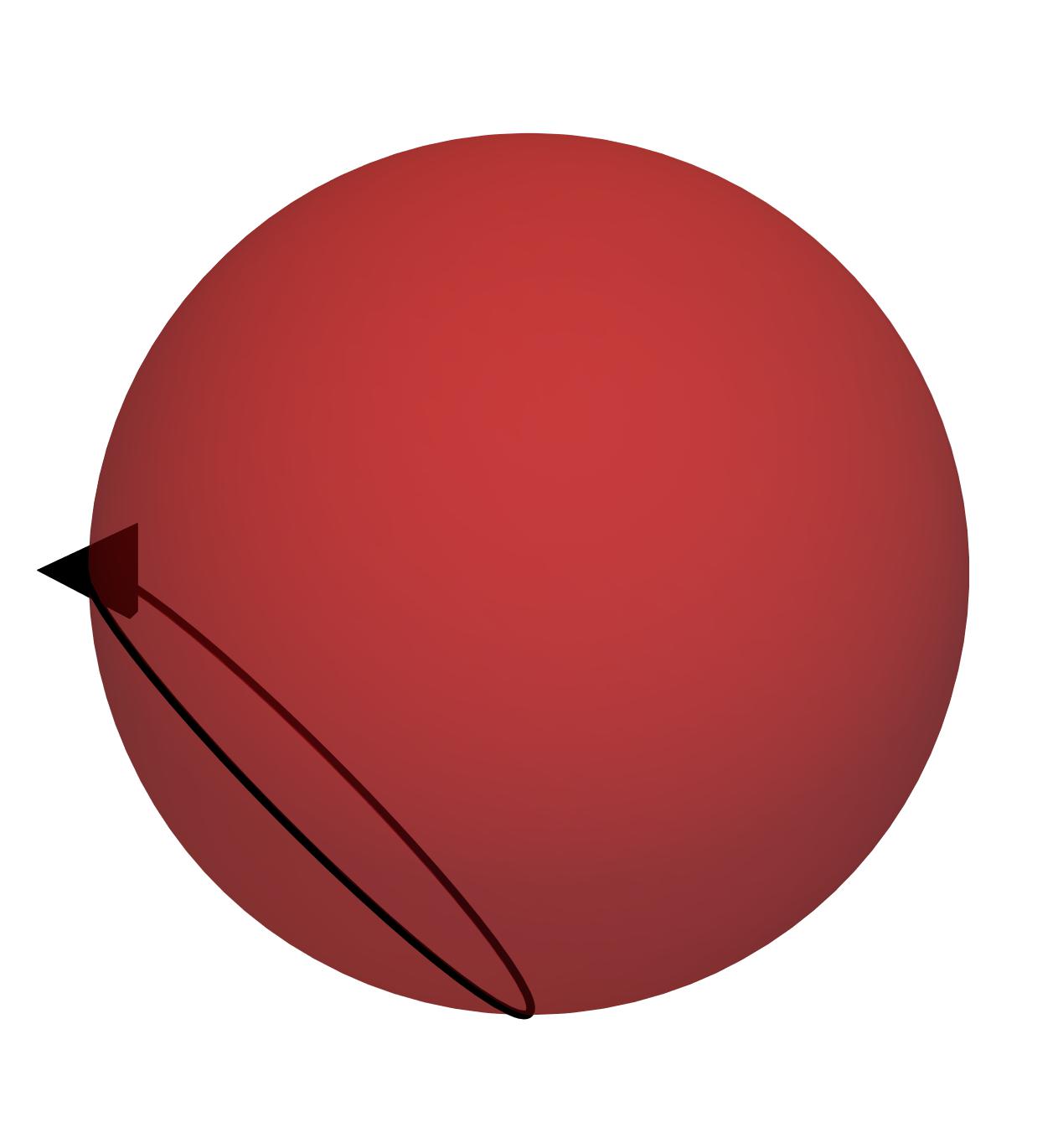}
\includegraphics[trim=50 70 50 80,clip,width=0.265\columnwidth]{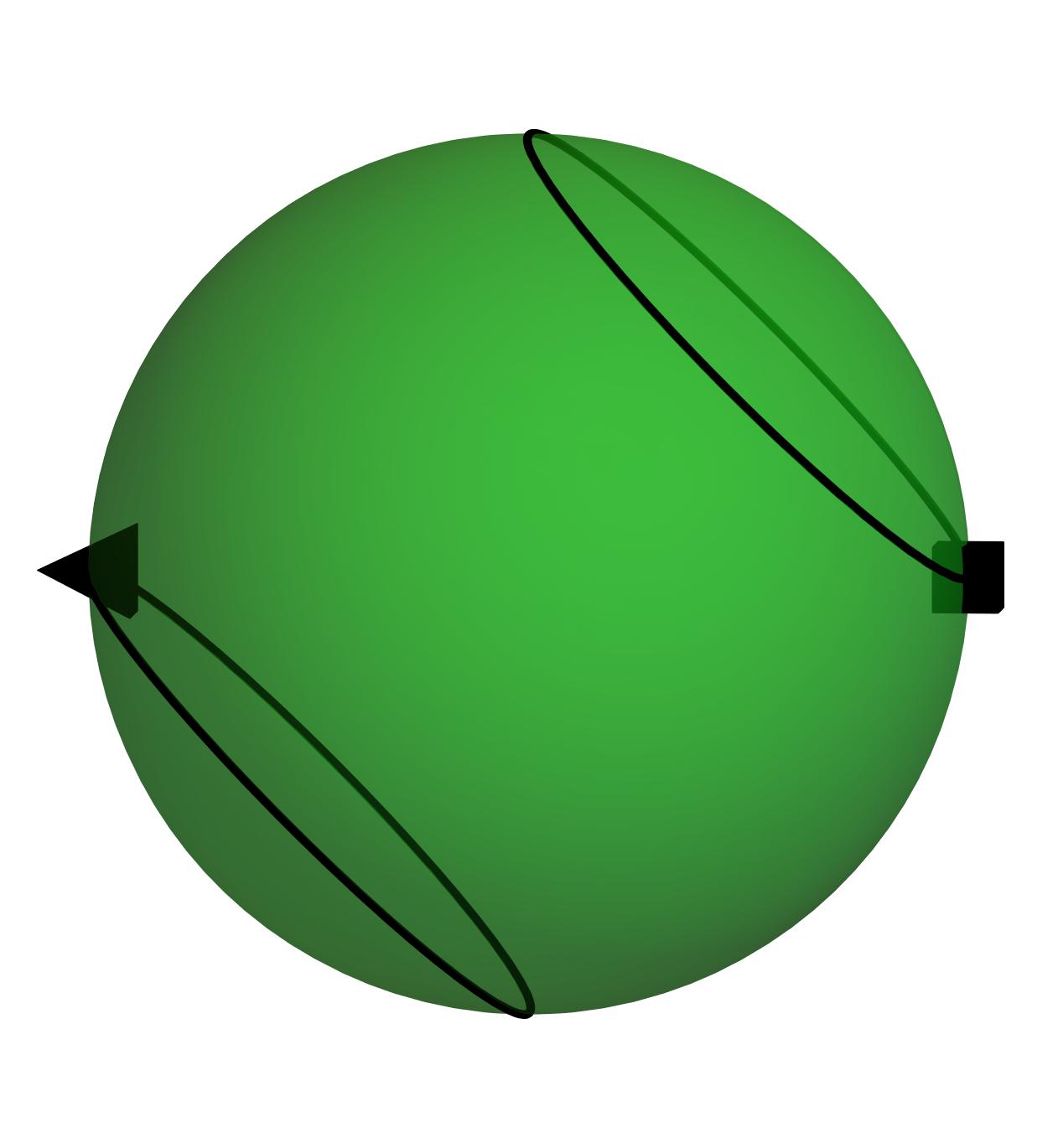}
\includegraphics[trim=50 70 50 80,clip,width=0.265\columnwidth]{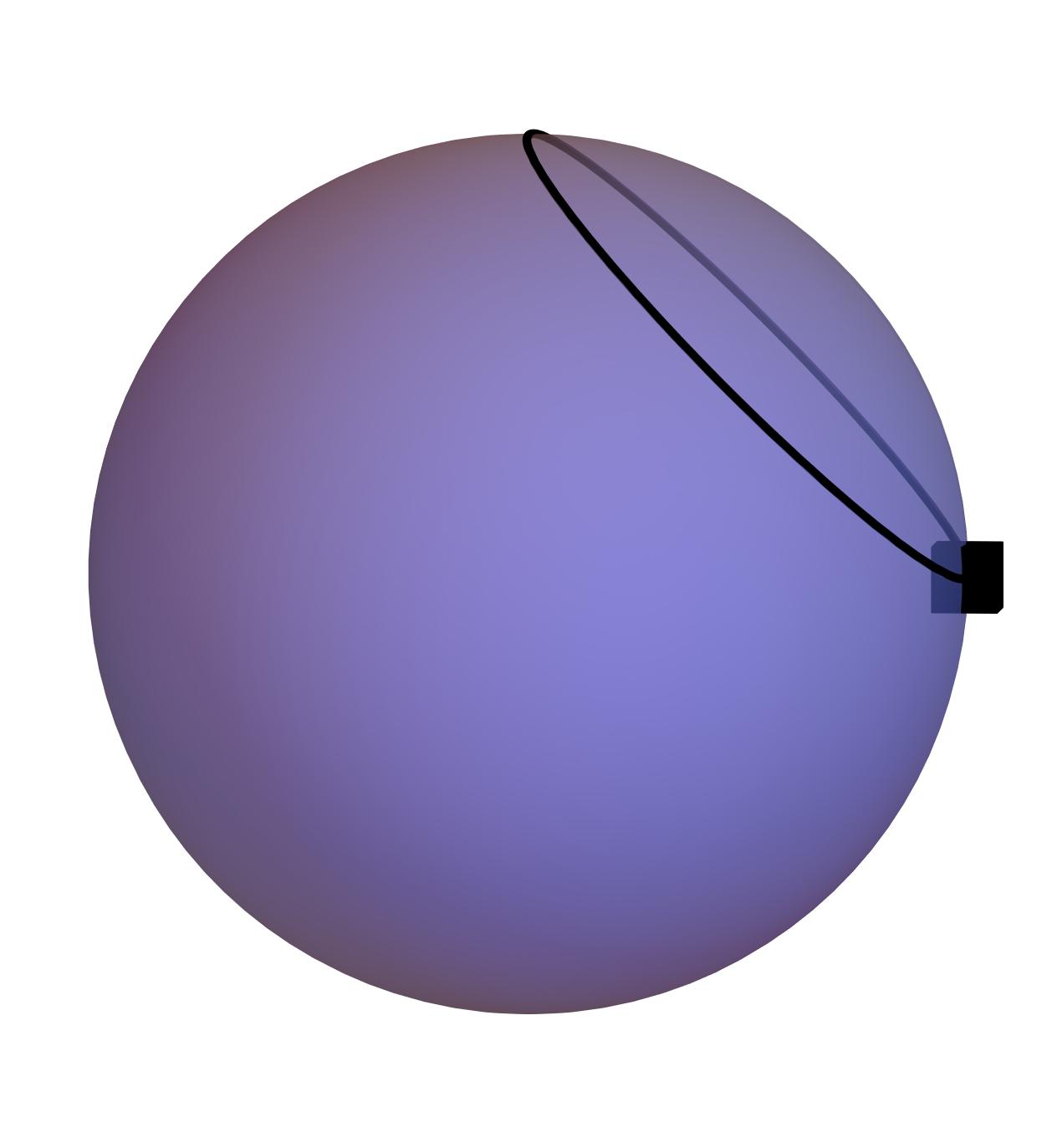}}
\boxed{\includegraphics[trim=50 70 50 80,clip,width=0.265\columnwidth]{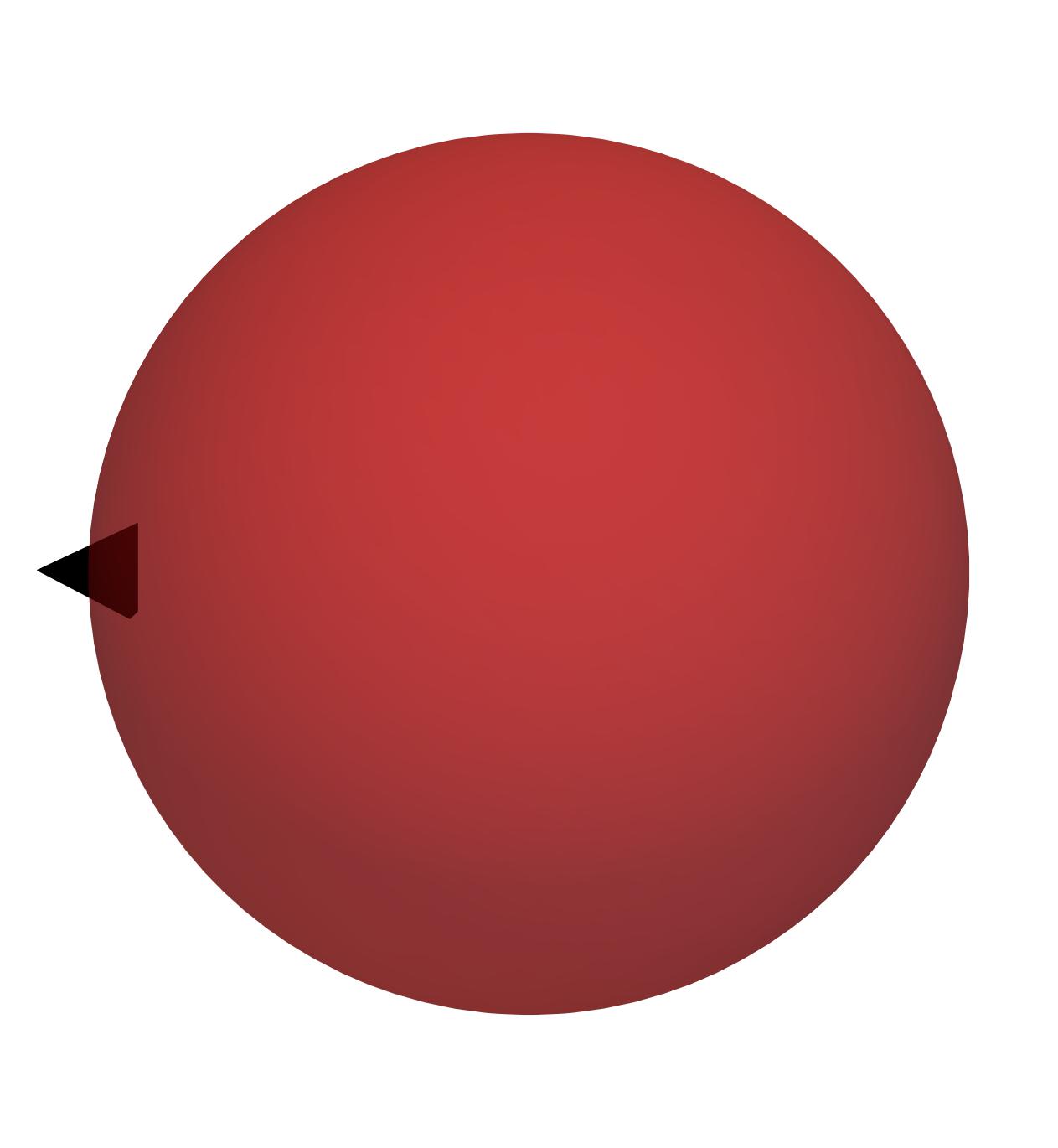}
\includegraphics[trim=50 70 50 80,clip,width=0.265\columnwidth]{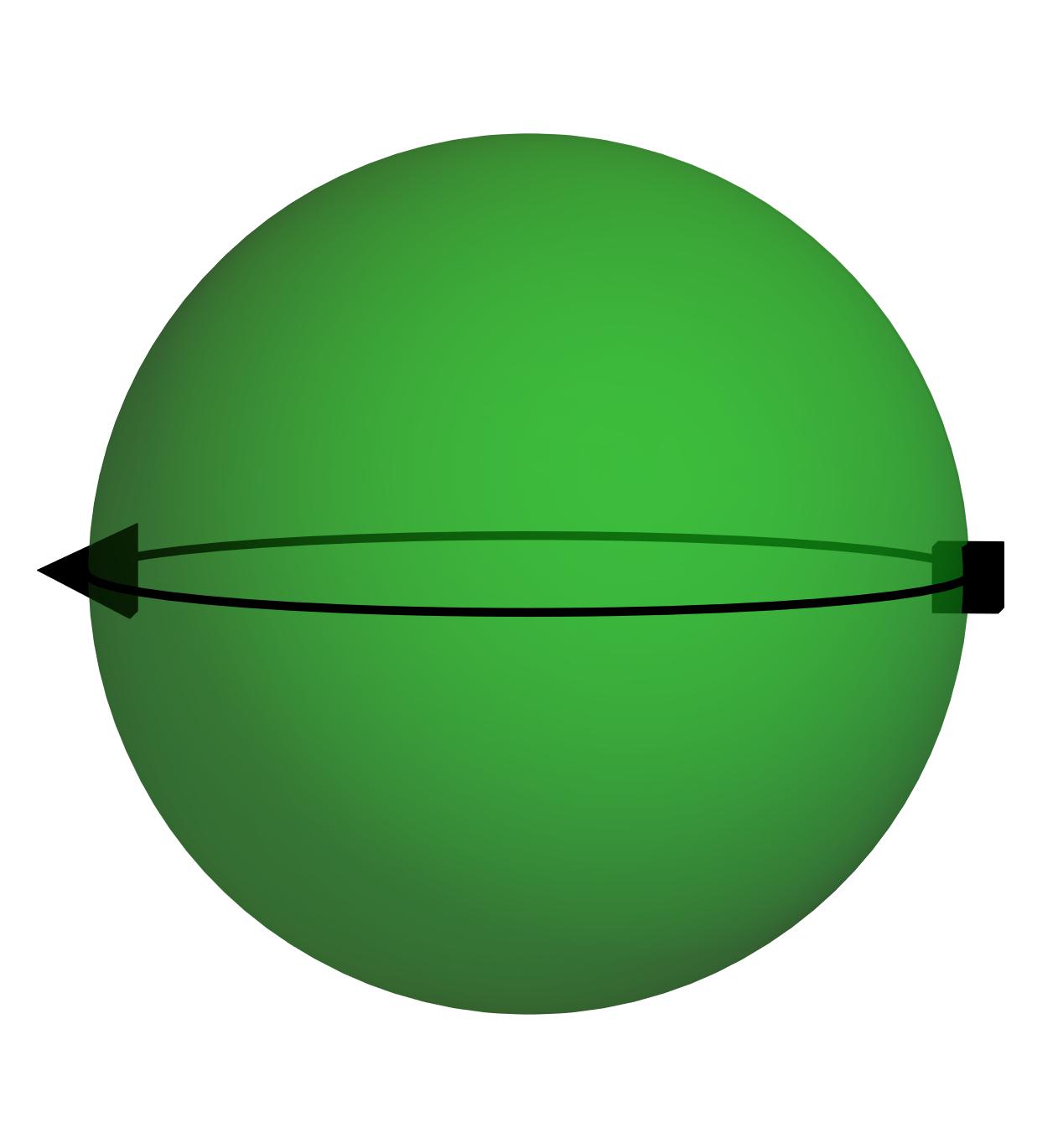}
\includegraphics[trim=50 70 50 80,clip,width=0.265\columnwidth]{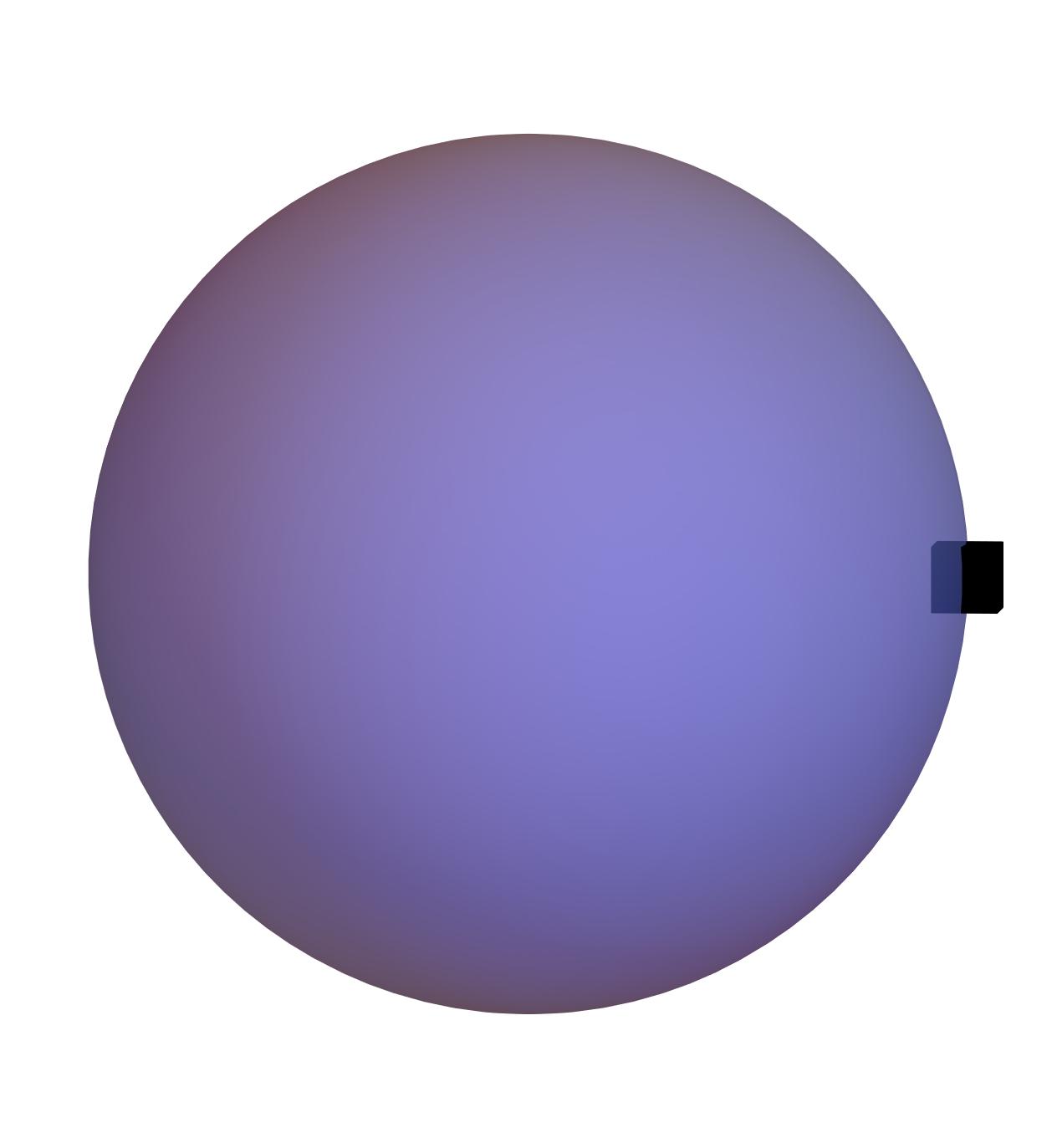}}
\caption{The homology classes of the Nexus type I triple point.
\cite{Chang_etal_2017} The first panel shows
non-contractible loops that do not touch any connecting points.
The second panel shows loops that touch only one connecting point.
The third panel shows a loop which touches two connecting points.
The loop drawing for this case follows the rule
in the bottom panel of Fig. \ref{fig:qbt_move_sketch}.}
\label{fig:homology_nexus_type1}
\end{figure}

With these basic steps in hand, we can enumerate all the 
non-trivial homological classes 
and they are shown in Fig. \ref{fig:homology_h8}.
There are three categories of non-trivial loops.
They are
\begin{enumerate}
\item loops involving only two connecting points,
they can be either on the left-middle sphere pair, or 
middle-right sphere pair.
\item Loops involving all the connecting points,
the two connecting points on the left and right spheres have to 
be joined, while on the middle sphere we have the two choices 
shown in Fig. \ref{fig:homology_h8}.
\item Loops on the same sphere that enclose the connecting
points.
\end{enumerate}

For the case of $H^3$, the generalized enclosing surface 
is shown
in second panel of Fig. \ref{fig:H3_homology}. For this case
there is only one possible non-contractible loop in the middle
sphere. This captures the band topology of $H^3$ and shows its
distinction from $H^8$ (and other cases).
From the above discussions, we can immediately
conclude that the $\Lambda^8$ triple point and 
two different $\Lambda^3$
and $\Tilde{\Lambda}^3$ triple points inside
the enclosing surface are not topologically different.
In our scheme, the distinction between different topological 
cases are categorized using the non-contractible loops 
or 1-cycles.
The number of distinct loops only depends on the number
(and kind) of the connecting points (Dirac-like, or possibly
QBT as in the examples to follow) on the enclosing
surface. Thus one cannot distinguish between pair of 
$\Lambda^3,\Tilde{\Lambda}^3$ triple points and a single
$\Lambda^8$ triple point which gives us a thumb rule
for composition of these triple point topological defects.

We end with an application of our scheme to recent Nexus
triple points discussed in the literature which have possible
material realizations. \cite{Chang_etal_2017}
For the type II nexus system as notated by Chang \emph{et al},
there are four line degeneracies
coming out of the triple point: one along the $z$-axis and
the rest three oriented at $\frac{2\pi}{3}$ angular separation
about the $z$-axis lying in high symmetry planes. 
See Eq. 2,3 in Ref. \onlinecite{Chang_etal_2017} for
the low-energy Hamiltonian, and Fig. 1 for the
line degeneracy structure.
This happens due to the presence of $\mathcal{C}_{3z}$
crystal symmetry.
\cite{Chang_etal_2017,Winkler_Singh_Soluyanov_2019}
In this case, the generalized domain for the surface 
enclosing the triple point degeneracy will have three spheres
connected to each other at the points 
where they intersect the line degeneracies.
The topology of this system can thus be similarly
understood using the homology classes as discussed above.
On this surface the loops are again of three main types
(diagram not show due to proliferation of non-contractible
loops): 
1) loop enclosing one connecting point, 
2) loop spanning two spheres, 
3) loop spanning through 3 spheres.
Even though these three types were also present in
the case of $H^8$, the count of each type is different 
which topologically distinguishes the two cases.

The case of type I as notated by Chang \emph{et al} is
worth noting. The generalized enclosing surface in this case
looks similar to that of $H^3$ (Fig. \ref{fig:H3_homology}).
However, the \ADEDITOKAY{classification}{characterization} of non-contractible loops is
different than the $H^3$ case. 
This is due to the degeneracies being QBT-like in this case. 
\cite{Chang_etal_2017} Thus while drawing the loops,
we have to follow the rule as shown in Fig.
\ref{fig:qbt_move_sketch}'s bottom panel. 
This allows for a new kind of non-contractible loop 
on the same sphere which goes through the
connecting point. We show the various possible loops in Fig. 
\ref{fig:homology_nexus_type1}.
This new kind of loops as in middle and bottom panel
of Fig. \ref{fig:homology_nexus_type1} are not possible when the
connecting points are Dirac-like because in that case
we are necessarily forced to go to the connected sphere
due to analyticity (Fig. \ref{fig:dirac_move_sketch}).
We remark here that Ref. \onlinecite{Chang_etal_2017}'s
statement that the line degeneracies are characterized by
a $2\pi$ Berry phase does not paint the full picture.
Such a characterization \emph{strictly} 
can only be applied to the non-contractible
loop shown on the top panel
of Fig. \ref{fig:homology_nexus_type1} and not in general.
Our scheme helps to make clear which loops have a topological
property based on analyticity. Once we have such loops in 
hand, we may compute familiar topological invariants
\cite{Heikkila_Volovik_2015} on the gapped ones among them.


\section{Conclusions and Discussions}
\label{sec:conclusion}

In summary, we laid a general scheme to describe the band
topology of the so-called Nexus triple point fermions. This
was based on an understanding of the analyticity properties
near (line-)degeneracies which are an integral part of the 
Nexus band structure. This scheme is built on the insight 
gained in Ref. \onlinecite{Das_Pujari_2019} where we could see the
analyticity properties near a line degeneracy explicitly.
The discussion started with the known cases of Dirac and
QBT bands in $2d$ in Sec. \ref{sec:2d_analyticity}.
We use analyticity to define a generalized domain
where we go smoothly across the ``connecting" points at the
degeneracies (see Fig. \ref{fig:dirac_move_sketch}). 
We emphasise that in the original domain there
is a non-analyticity in the space of wavefunctions at a 
(non-accidental) degeneracy which is then considered as a 
topological defect. In the generalized domain, however,
this issue is not there.
For the $2d$ Nexus case, the generalized domain is a familiar
object -- Riemann surfaces associated with $z^{1/3}$ -- 
known from the study of complex analysis. However, the general
idea is applicable in any situation.
So we take this scheme to $3d$ and define generalized
domains for 3d Nexus triple points in Sec. \ref{sec:3d_analyticity}.

In analogy with Weyl points and Chern numbers on associated 
enclosing surfaces, we characterize the $3d$ Nexus points
by enclosing them in the generalized domain (see bottom panel of 
Fig. \ref{fig:h8_enclosing_surface}, \ref{fig:H3_homology}) in a
departure from existing literature. 
Sec. \ref{sec:classification} describes the triple point defect
topology in terms of non-contractible loops that can be drawn 
on this enclosing surface. These are the 1-cycle homology classes
of the generalized domain.
Different Nexus triple points have their unique data of these
1-cycle homology classes. This discrete set of data
gives the triple point its topological character, since they
will be stable to small deformations of the Hamiltonian.
We reiterate here again that this way of describing the topology
is actually more general (e.g. we can enclose multiple
Nexus points, etc.), however, we have principally 
concerned ourselves with single Nexus triple points. 
Line-degeneracies on the other hand are characterizable by using
topological invariants defined on the gapped loops around them.
\cite{Heikkila_Volovik_2015} Our enclosing scheme is finally
applied to examples of Nexus triple point in the literature
which has possible material realizations, \cite{Zhu_etal_2016,Chang_etal_2017}
whereas only the topology of gapped enclosing loops
around the line degeneracies and their evolution
across the triple point had previously been discussed.
\cite{Chang_etal_2017,Zhu_etal_2016,Winkler_Singh_Soluyanov_2019}

\subsection{Surface Fermi Arcs}

This final result of our paper provides an answer
to the question of Fermi arc protection in Nexus
systems that was raised by Ref. \onlinecite{Chang_etal_2017}.
We restrict our discussion to the zero or weak spin-orbit coupled case
for simplicity as in Type I of Ref. \onlinecite{Chang_etal_2017}.
However, there can be more general situations with 
multiple triple points in a multi-band system in the presence of
spin orbit coupling \cite{Zhu_etal_2016} where the Fermi arcs
can be more elaborate whose exact structure depends on the actual details
of the bandstructure.
For the restricted case, 
since the Nexus triple points are topological in nature,
therefore the associated surface arcs will be protected
and will necessarily go through the surface projections of the
Nexus triple point. 
We can already conclude that there will at least be two protected
Fermi arcs because of the following:
In case of a Weyl system, we know that the total Chern number of
filled bands on $2d$ cross-sections changes
across the Weyl point which leads to the existence of the Fermi
arcs (see Sec. II-C-1 of Ref.
\onlinecite{Armitage_Mele_Vishwanath_review_2018}).
For a Nexus system with the Nexus points assumed to
lie close to the Fermi level,
there will be two filled $2d$ bands on generic cross-sections
on one side, while there will be a single filled
$2d$ band on the other side as already seen in Sec.
\ref{sec:3d_analyticity}.
Now, the total Chern number of the filled bands
on either side is zero. Thus, there cannot be a non-zero Hall
conductance. However, the two filled bands have a
non-zero chiral winding number (Sec. \ref{subsec:nexus_anal}), while the single filled band does
not have any such winding. Due to this winding number change across
Nexus points, there will \emph{at least} be two counter-propagating
zero modes on the $2d$ boundary to ensure that the Hall 
conductance is zero, thereby leading to two surface Fermi arcs on
the $3d$ boundary. Presence of two surface arcs has been seen in 
numerics. \cite{Zhu_etal_2016,Chang_etal_2017}
An interesting question remains as to the effect of
the chiral winding number on the charge of these edge modes. We
conjecture that the charge may not be unity for higher
chiral winding numbers.

\subsection{Outlook}

We end with some discussion on
the conceptual issues that still remain to be
understood. One thing that we have puzzled over is whether there
exists a Chern number like description of the Nexus triple point
topology by making use of the Berry connection/curvature
technology, in spite of the absence of a gapped enclosing surface
which motivated the entire line of reasoning in this paper. 
Instead of thinking as a single analytic ``band" defined on the
generalized domain which gave us our homological \ADEDITOKAY{classification}{characterization}
scheme, if we think of three bands on the 
conventional domain, then the Dirac points are like monopoles on
the enclosing surface. The associated Berry curvature will
thus diverge at the degeneracy points on the sphere. So the
integral of the Berry curvature over the sphere is not guaranteed
to be well-defined. Could there still be a finite piece in this 
integral which may capture the underlying topological nature?

Another approach could instead be to consider a non-Abelian
characterization. In fact, this approach can be implemented
for the $2d$ example $H$ introduced in 
Sec. \ref{sec:2d_analyticity}. \cite{pujari_youtube_2019}
A similar implementation in $3d$ is not yet clear to us, but we may
anticipate a matrix of topological charges instead of a single
scalar charge. Finally, some other mathematical machinery might
be useful that we don't anticipate yet.

We end with some final thoughts on connecting the homological
loops to possible experimental observable. As mentioned before,
the topological character of degeneracies in the bulk
have profound effects on the surface states. 
Thus for the case of the Nexus triple point, 
we may specifically ask how the homological loop 
classes identified  in this paper -- especially the ones 
which live on multiple
spheres -- affect the surface states. Each homological
class may leave its own distinct imprint on the surface states
which can perhaps be identified in experiments or simulations.
Of course, the effect of electron-electron
interactions \cite{Sim_etal_2019} or disorder on Nexus fermions 
are yet to be fully explored.

\begin{acknowledgements}
SP acknowledges financial
support from IRCC, IIT Bombay (17IRCCSG011) and
SERB, DST, India (SRG/2019/001419). AD thanks NSF-DMR-1306897 and 
NSF-DMR-1611161 for financial support. 
AD also thanks Weizmann Institute of Science, Israel Deans
fellowship and Israel planning and budgeting committee for
financial support. This research was supported in part by the
International Centre for Theoretical Sciences (ICTS) during a
visit for participating in the program -  Novel phases of quantum 
matter (Code: ICTS/topmatter2019/12).
\end{acknowledgements}

\bibliographystyle{apsrev}
\bibliography{3dAnal}

\begin{thebibliography}{39}
\bibitem[{foo()}]{footnote_winding}
\bibinfo{note}{Windings numbers are more complete in a sense, e.g. they capture
  better how a single quadratic band touching splits into four linear band
  touchings under $C_3$ symmetry in bilayer graphene.}
\end{thebibliography}

\end{document}